\documentclass[12pt,tightenlines,nofootinbib]{revtex4-2}
\usepackage{amsfonts,amsmath,amssymb}
\usepackage{enumitem}
\usepackage{txfonts}
\usepackage{graphicx,subfigure,wrapfig}
\usepackage{color,xcolor}
\usepackage{lipsum,booktabs}
\usepackage{floatrow,float}
\usepackage{epstopdf}
\usepackage{hyperref}
\usepackage[normalem]{ulem}
\usepackage{wasysym}

\usepackage{lineno}
\begin{document}

\title{Altered patterning of neural activity in a tauopathy mouse model}
\author{C. Hoffman$^1$, J. Cheng$^2$, R. Morales$^1$, D. Ji$^2$, Y. Dabaghian$^1$}
\affiliation{$^1$Department of Neurology, The University of Texas McGovern Medical School, 6431 Fannin St, 
	Houston, TX 77030\\
$^2$Department of Neuroscience, Baylor College of Medicine, Houston, TX 77030,\\
$^{*}$e-mail: Yuri.A.Dabaghian@uth.tmc.edu}
% other emails
%\vspace{17 mm}
\date{\today}

\begin{abstract}
	Alzheimer's disease (AD) is a complex neurodegenerative condition that manifests at multiple levels
	and involves a spectrum of abnormalities ranging from the cellular to cognitive. Here, we investigate
	the impact of AD-related tau-pathology on hippocampal circuits in mice engaged in spatial navigation,
	and study changes of neuronal firing and dynamics of extracellular fields. While most studies are based
	on analyzing instantaneous or time-averaged characteristics of neuronal activity, we focus on intermediate
	timescales---spike trains and waveforms of oscillatory potentials, which we consider as single entities.
	We find that, in healthy mice, spike arrangements and wave patterns (series of crests or troughs) are
	coupled to the animal's location, speed, and acceleration. In contrast, in tau-mice, neural activity is
	structurally disarrayed: brainwave cadence is detached from locomotion, spatial selectivity	is lost, the
	spike flow is scrambled. Importantly, these alterations start early	and accumulate with age, which exposes
	progressive disinvolvement the hippocampus circuit in spatial navigation. These features highlight
	qualitatively different neurodynamics than the ones provided by	conventional analyses, and are more salient,
	thus revealing a new level of the hippocampal circuit disruptions.
	\newline
	\newline
	\newline
	%\vspace{17pt}
		\textbf{Significance}. We expose differences in WT and tau brains, emerging at the circuit level, using
		a novel, morphological perspective on neural activity. This approach allows identifying qualitative
		changes in spiking patterns and in extracellular field oscillations, that are not discernible through
		traditional time-localized or time-averaged analyses. In particular, analyses of activity patterns 
		facilitates detection of neurodegenerative deviations, conspicuously linking their effects to behavior
		and locomotion, thus opening a new venue for understanding how the architecture of neural activity 
		shifts from normal to pathological.
\end{abstract}
	
\maketitle
\newpage

\section{Introduction}
\label{sec:intro}

Alzheimer's Disease (AD) is a devastating neurodegenerative condition that is extensively studied at multiple
levels, from molecular, to cellular, to organismal \cite{DeTure,Perl,Wenk}. It is particularly challenging to
understand how all these levels connect and to identify which functional intermediaries link neural pathologies
to cognitive deterioration. For instance, it is believed that $A\beta$-plaques and neurofibrillary tangles
disrupt the synergistic coordination of circuit dynamics \cite{Targa,Bachmann}. However, their subsequent
contribution to the observed behavioral symptoms of AD remain unclear \cite{Alzheimer,Glenner,Oddo,Selkoe2}.
Many studies of AD-related changes of spiking activity and synchronized extracellular field dynamics reveal
disturbances, including alteration of neuronal firing rates \cite{Martinsson,Frere}, slowing of brainwave
rhythms \cite{Rae-Grant,Liddell}, reductions in signal complicacy \cite{Jelles}, and increased epileptiform
activity \cite{Palop,Born,Palop2,Pandis,Vossel,Scarmeas}. Despite these efforts, systemic insight into the
circuit-level impact of AD pathologies is still lacking due to the prohibitive complexity of the network
dynamics, sheer amount of interconnected elements, and intricacy of interactions which render detailed data
analyses and realistic modeling nearly intractable.

On the other hand, this very intricateness begets simplification by bringing forth statistical aspects of
circuit activity. As it turns out, large pools of spike trains and waveforms exhibit universal statistical
properties that have generic, mathematical origins, independent from the underlying physiological mechanisms.
The ways in which these statistics unfold in time open a new venue for examining circuit dynamics differences 
between normal and pathological brains. 

%%%%%%%%%%%%%%%%%%%%%%%%%%%%%%%%%%%%%%%

\begin{figure}[H]
	\centering
	\includegraphics[scale=.9]{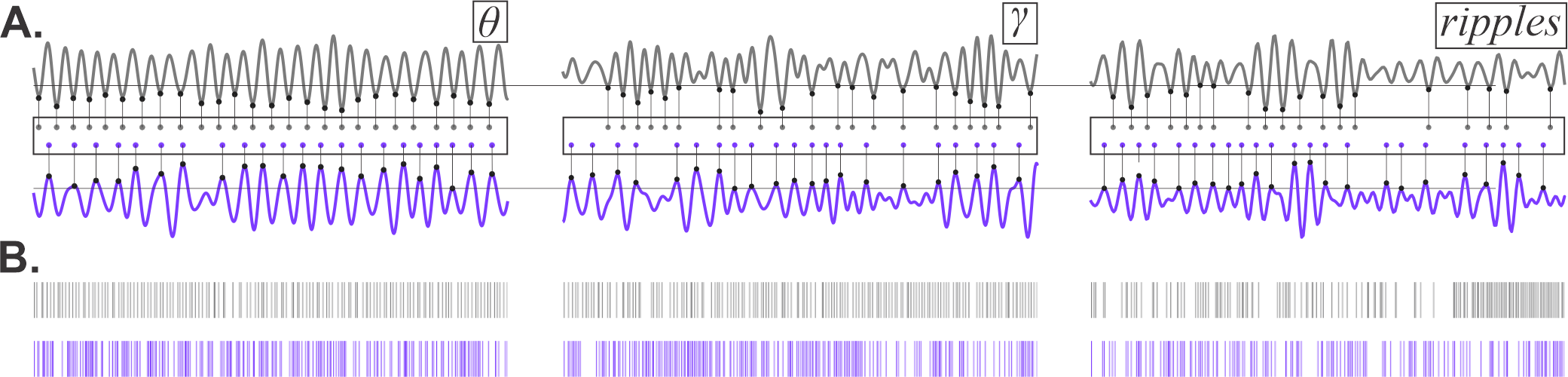} 
	\caption{{\footnotesize
			\textbf{Patterns of neural activity in wild type (WT) and and mice affected by brain tauopathy}. 
			\textbf{A}. LFP waves recorded in WT (gray trace) and tau (purple trace) mice, with their
			significant peaks and troughs marked by dots, boxed. Differences between the resulting patterns
			are apparent, although the waves are scaled to the same frequencies and amplitudes. The shapes of
			the	waves are easily recognizable, $e.g.$, the ones on the left, marked ``$\theta$," correspond
			to the $\theta$-waves ($4-12$ Hz). Would it be surprising if they were marked as ``$\gamma$" ($30-
			80$ Hz) or ``ripples" ($150-250$ Hz)? How surprising? Also note that tauopathic $\theta$-waveforms
			are somewhat less regular than the WT $\theta$-waveforms, whereas WTs' $\gamma$- and ripple
			waveforms appear more cluttered than in tau-mice. Could these differences be attributed to
			random fluctuations of the waves' shapes or should they be taken as signatures of qualitative
			changes? How to capture the morphological differences mathematically?
			\textbf{B}. Spike trains recorded from WT (gray) and tau (purple) interneurons also demonstrate
			different patterns, suggesting that tau-pathologies alter the flow of information exchange.
			%These differences can be objectively quantified, linked to behavior and used for assessing the
			%effects of tau-pathologies in the hippocampal network. 
	}}
	\label{fig:waves}
\end{figure}

%%%%%%%%%%%%%%%%%%%%%%%%%%%%%%%%%%
In this study, we consider neuronal spike trains and waveforms of locally recorded extracellular field
potentials (LFP) as basic observables characterizing circuit dynamics. We analyze these temporally extended
entities without averaging, evaluating their instantaneous behavior, or decomposing into simpler constituents. 
Instead, we use certain morphological properties of data segments to obtain their integral descriptions
at structural level, which opens a new narrative and creates new perspectives on the underlying physiological
phenomena. As an illustration of our approach, consider the brainwaves illustrated on Fig.~\ref{fig:waves}A,
which are scaled to same mean amplitude and frequency range. Yet, they are visibly different: one could argue,
$e.g.$, that top waves are more regular, with stereotypical undulatory shapes, or perhaps more ordered. 
Similarly, the spike trains shown on the top of Fig.~\ref{fig:waves}B are less cluttered, less haphazard and
more uniform than their bottom counterparts. Surprisingly, these intuitive distinctions can be captured formally
and used for quantifying neurodynamics \cite{Hoffman}. In the following, we apply these methods to study 
hippocampal activity in a mouse model of tau pathology and extracted a number of constitutive, pathological
deviations of their circuit activity.

%Methodologically, this approach bridges the gap between nigh, instantaneous analyses of neuronal activity,
%and large-scale, time-averaged descriptions.

%The paper is organized as follows. In Sec.~\ref{sec:met} we outline formal mathematical approaches that allow
%such evaluations of spike patternings and LFP shapes. In Sec.~\ref{sec:res}, we use these methods for describing
%hippocampal neurodynamics in mice of both phenotypes. In Sec.~\ref{sec:disc}, we discuss the implications of our
%results. Mathematical details are provided in Sec.~\ref{sec:math}. 
%provide underpinnings for such notions and th
%``uncharacteristic,"
%\textcolor{blue}{allow qualitative reasoning about the brain waves’ shapes associated with AD?}

\section{Methods}
\label{sec:met}

\textbf{Experimental procedures}. Spike and LFP data were recorded from the hippocampal CA1 area of healthy
wild type (WT) and transgenic mouse models of tau pathology (rTg4510). rTg4510 mice which develop tau 
neurofibrillary tangles and neuronal loss similar to those observed in human tauopathies \cite{Ramsden,SantaCruz}.
The animals were trained in a familiar room to run back and forth on about $2$ m long rectangular track for
food reward. The daily recording procedure consisted of two $15$-minute-long running sessions, followed by 
$15$-minute sleep breaks. The LFP data was sampled at $2$ kHz rate and the animals' positions were sampled at
$33$ Hz with a resolution of approximately $\pm 0.2$ cm, by tracking two head-mounted color diodes. Further
details on the surgery, tetrode recordings and
other procedures can be found in \cite{ChengJi,Ciupek}. %The recording was repeated $3–10$ days.

\textbf{Computational procedures} are based on applying two complementary measures.

\begin{figure}[H]
	\centering
	\includegraphics[scale=.81]{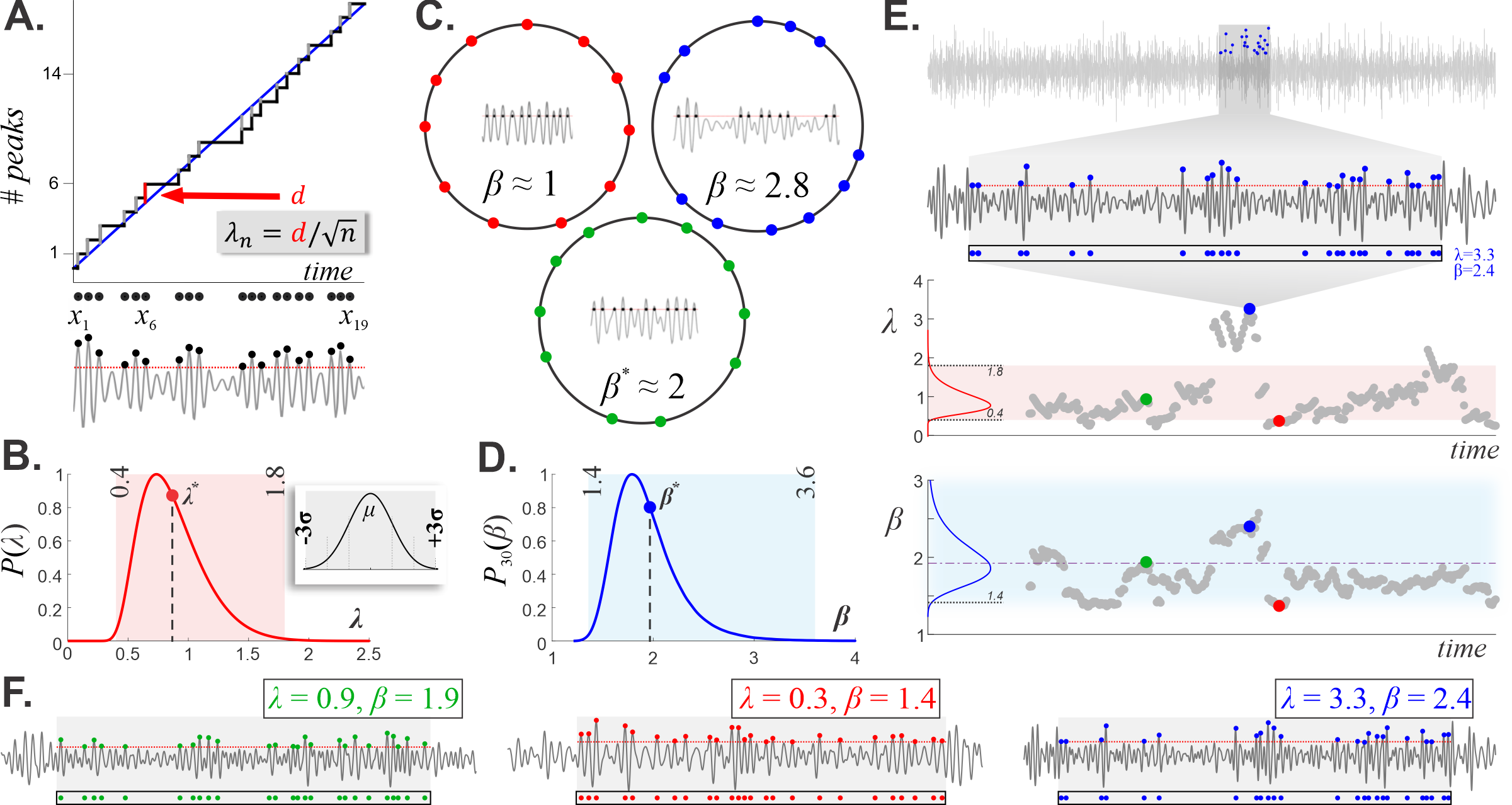}
	\caption{{\footnotesize 
			\textbf{Pattern statistics}. 
			\textbf{A}. Given a series of values, $X=\{x_1,x_2,...,x_n\}$, $e.g.$, a series of $n=19$ wave crests
			shown at the bottom, we build its \textit{empirical distribution}---a staircase that makes a unit
			step up at each consecutive $x_i$. The slope of the staircase is defined by the mean trend (blue
			line), $e.g.$ about eight $\theta$-peaks are expected every second. The $\lambda$-score of $X$ is the
			greatest deviation of its points from this trend, normalized by the sample size. 
			\textbf{B}. The distribution of $\lambda$-scores, $P(\lambda)$, is universal and defines the
			impartial probability of a given pattern's appearance. On average, $\lambda$-scores are close to
			$\lambda^{\ast}\approx 0.87$; patterns that closely follow the prescribed behavior produce small
			$\lambda$s, and those that deviate significantly from the mean trend have large $\lambda$-scores.
			About $99.7\%$ of patterns score between $\lambda^{-}=0.4$ and $\lambda^{+}=1.8$ (pink stripe) and
			are hence ``typical", whereas patterns with small ($\lambda<0.4$) or large ($\lambda>1.8$) 
			$\lambda$-scores are ``atypical". The selection of bounds is motivated by analogy with the Gaussian
			distribution (top panel), in which typical values are those that fall	closer than	$3\sigma$ from
			the mean (gray stripe, $99.7\%$ of cases) and the atypical ones lay farther out ($0.3\%$).
			%\footnote{Throughout the text, terminological definitions and highlights are given in italics}.
			\textbf{C}. Placing $n$ elements of $X$ around a circle, summing the squared arc lengths, and
			normalizing the result by the circumference yields $\beta(X)$. For periodic arrangements (red
			points) $\beta$ is small, $\beta\approx1$, for ``clustered" patterns (blue points) $\beta$ is large
			(up to $\beta\approx n$), and for generic layouts (green points) $\beta\approx2$---a surprising 
			statistical universality. 
			\textbf{D}. The distribution of $\beta$-values for a pattern, $X$, of length $n=30$: similar to
			$\lambda$-scores, typical $\beta$ values fall between specific bounds $\beta^{-}_{n=30}=1.4$ and
			$\beta^{+}_{n=30}=3.6$. The general form of $\beta$-distributions depends on the pattern's size, 
			as illustrated on Fig.~\ref{fig:3D}D. 
			\textbf{E}. Shown are $\lambda$ (top) and $\beta$ (bottom) dynamics, evaluated by ``sliding" a
			$25$-peak window along a $\gamma$-wave. The pink stripe represents the typical range of $\lambda$s
			as on panel B (laid horizontally), and the blue stripe is the same as on panel D.
			\textbf{F}. An example of a ``statistically mundane" pattern (green point on panel E, $\lambda=0.9$
			and $\beta=1.9$) and patterns with small or large values of $\lambda$ and $\beta$ (red and blue
			points on E, respectively).
	}}
	\label{fig:stoch}
\end{figure}

\textit{$\lambda$-score}, introduced by A. Kolmogorov, quantifies the ``randomness" or ``haphazardness"
of patterns through their deviation from an expected mean trend (Fig.~\ref{fig:stoch}A). Remarkably, these
deviations exhibit regular statistical behaviors, described by a universal probability distribution,
$P(\lambda)$ (Fig.~\ref{fig:stoch}B, \cite{Kolm,Arn1,Arn2,Arn3}). According to its structure, most patterns
produce $\lambda$-scores confined between certain limits, $e.g.$, about $99.7\%$ of patterns have
$\lambda$-scores between $\lambda^{-}=0.4$ and $\lambda^{+}=1.8$, and only $0.3\%$ fall outside of these
bounds. Thus, the patterns with $\lambda$-scores within that range can be qualified as ``stochastically
typical," whereas those that fall outside of it are ``atypical." 

In practice, the mean trend of recurring neurophysiological activity is often easy to estimate, $e.g.$, one can
evaluate the mean rate of LFP oscillations by averaging the number of crests or troughs over sufficiently long
periods---such as a running session. However, these trends are known to change with the animal's physiological
state. For instance, the LFP's overall cadence alters between active movement and quiescence 
\cite{Murthy,Rickert,Baker} (Fig.~\ref{fig:supp2}). Correspondingly, the ``reference point" for computing the
patterns' statistics also changes. Taking these alterations into account allows for \textit{contextual} 
$\lambda$-scores that produce more nuanced descriptions of circuit dynamics, discussed below.

\textit{$\beta$-score} provides an alternative measure of stochasticity that emphasizes patterns' orderliness
(Fig.~\ref{fig:stoch}C). As shown in \cite{Arn1,Arn2,Arn3}, nearly periodic, ordered patterns of $n$ elements
produce smaller $\beta$-values (the minimal value, $\beta=1$, is produced by a strictly regular, periodic
pattern). Clustering patterns yield larger $\beta$-values: the maximum, $\beta=n$, is reached when all elements
lump together. As it turns out, impartially scattered patterns produce $\beta$-scores close to $2$---as a matter
of another surprising universality \cite{Hoffman,Arn1}. 

Curiously, the shapes of the $\beta$-distributions, $P_{n}(\beta)$, albeit $n$-dependent, are similar to the
shape of Kolmogorov distribution, $P(\lambda)$, in one important aspect: they all have a dominant ``hump" and
rapidly decaying tails (Fig.~\ref{fig:stoch}D). Moreover, all $\beta$-distributions have similar modes and
means, that converge to $\beta^{\ast}\approx2$ as the sample size increases (Fig.~\ref{fig:3D}). One can
therefore specify limits, $\beta_{n}^{-}$ and $\beta_{n}^{+}$, that confine the ``typical" $\beta$-scores and
exclude the ``atypical" ones. For instance, $\beta_{30}^{-}\approx1.4$ and $\beta_{30}^{+}\approx3.6$, confines
$99.7\%$ of patterns containing $n=30$ elements (Fig.~\ref{fig:stoch}D). 
%%%%%%%%%%%%%%%%%%%%%%%%%%%%%%%%%%

%%%%%%%%%%%%%%%%%%%%%%%%%%%%%%%%%%
\textit{Dynamics of stochasticity} can be computed by evaluating the $\lambda$- and $\beta$-scores for local
data segments, $e.g.$, for waveforms or spike trains contained in a time window of width $L$, centered at time
$t$. As the window shifts forward in time, this ``snapshot" of the data evolves, yielding time-dependent 
haphazardness, $\lambda(t)$, and orderliness, $\beta(t)$, that describe pattern changes. We emphasize here
that circuit dynamics can drive the brain waves into assuming highly improbable waveforms, characterized by
very low probabilities of accidental appearance. For example, periodic patterns are too orderly to happen
by chance, but they can be enforced by limit cycles in the network's phase space. In the following,
we detect qualitative changes in the hippocampal dynamics by identifying atypical network patterns. 

In order to construct $\lambda(t)$ and $\beta(t)$ dynamics empirically, one can either use a time window of 
fixed width, shifting it by a pre-specified time step, or adapt the width at each step, so that each window
contains a pre-specified number of spikes or LFP peaks (Fig.~\ref{fig:FixedN}). Numerically, these algorithms
differ only marginally, but they catch different aspects of the patterns' statistics. Intuitively, the first
method captures changes in discretized ``physical" time, and the second method follows the alterations relative
to the network's own tempo of activity. In the following, we analyze the dynamics of LFP patterns using the 
latter approach, to highlight pattern changes with respect to the hippocampal network's own ebb and flow.

%\newpage
%\clearpage
\section{Results}
\label{sec:res}
%%%%%%%%%%%%%%%%%%%%%%%%%%%%%%%%%%%%%%%

%\begin{figure}[h]
\begin{wrapfigure}{c}{0.5\textwidth}
	\centering
	\vspace*{-10pt}
	\includegraphics[scale=0.93]{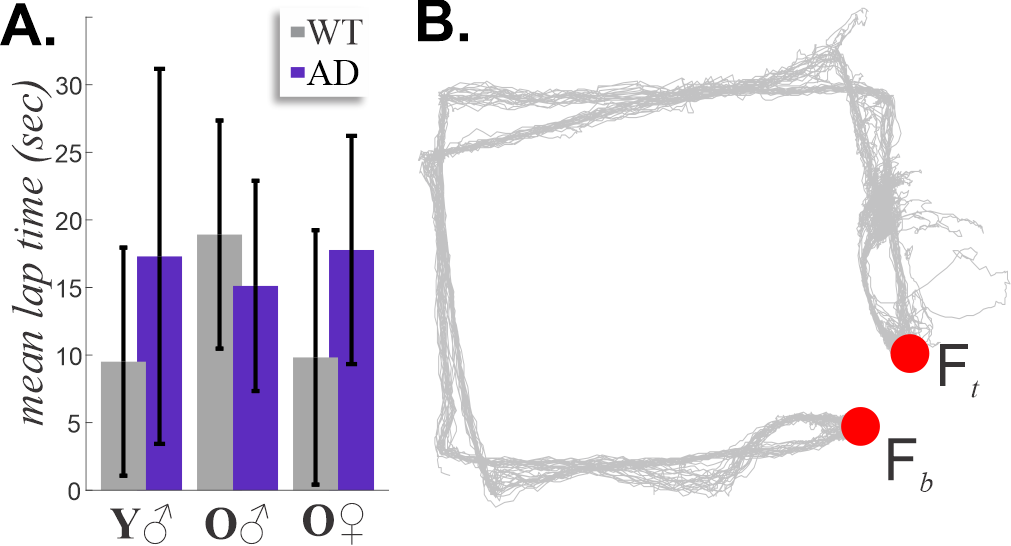}
	\caption{{\footnotesize
			\textbf{Locomotion}. 
			\textbf{A}. No significant difference in the times of lapping from one food well to the	other,
			for	all age groups and phenotypes ($p=0.42$). A typical lap takes young WT males ($Y^w_{\male}$)
			$10$ sec, young tau-males ($Y^{\tau}_{\male}$) $17$ sec, old WT males	($O^w_{\male}$) $19$
			sec, old  tau-males ($O^{\tau}_{\male}$) $15$ sec, old WT females ($O^w_{\female}$) $10$ secs,
			and old tau-females ($O^{\tau}_{\female}$) $18$ sec. 
			\textbf{B}. The gray lines trace the mouse's runs between top and bottom food wells, $F_b$ and
			$F_t$ (red dots), during one recording session. Behaviorally, inbound trajectories (horizontally
			aligned quadrilateral) and outbound trajectories (skewed quadrilateral) are different, $e.g.$, the
			animals	tend to turn slowly around the top right corner and dash to $F_t$, while $F_b$ is 
			approached more leisurely.
		}}
	\label{fig:track}
\end{wrapfigure}
%\end{figure}

%%%%%%%%%%%%%%%%%%%%%%%%%%%%%%%%%%
We used two groups of mice: ``young" ($Y$) and ``old" ($O$) animals, $2.4-3.8$ and $7-9$ months of age
respectively, and studied the patterns of their neuronal spiking and principal brain waves. 
There was significant histpathological difference between old WT and tau hippocampi, but not in young mice, 
including cortical thinning in old tau mice despite immunohistological confirmation of abundant abnormal tau 
protein in both young and old tau brains (for details see \cite{Ciupek}). On average,
it took the mice $15$ seconds to traverse the track (Fig.~\ref{fig:track}A), which included lapsing over the
straight segments, turning around the junctions, and pausing at the food wells (Fig.~\ref{fig:track}B).
To capture the dynamics of circuit activity in the context of ongoing behavioral and physiological states,
we studied patterns that lasted about $L\approx2$ sec (thrice shorter than in \cite{Hoffman} for improved
temporal resolution), which are long enough to produce stable scoring of waveforms and spike trains. 
%\subsection{Theta rhythm} 
%\label{sec:theta}

\textbf{$\theta$-rhythm} ($4-12$ Hz) plays major roles in spatial memory, cognition, movement and other
phenomena \cite{Burgess,BuzTheta1,BuzTheta2,BuzTheta3,Cacucci,Itskov,Osbert}. AD-induced alterations of
$\theta$-rhythmicity were reported in many studies, based on analyzing instantaneous and averaged parameters
of neural activity \cite{Jelic,Bennys}. We hence inquired whether tau-pathologies also affect the patterning
of $\theta$-waveforms, to which end we evaluated their haphazardness ($\lambda_\theta$) and orderliness
($\beta_\theta$) scores in both phenotypes, and followed their dynamics. 

\begin{figure}[H]
	%\begin{wrapfigure}{c}{0.7\textwidth}
	\centering
	\includegraphics[scale=0.8]{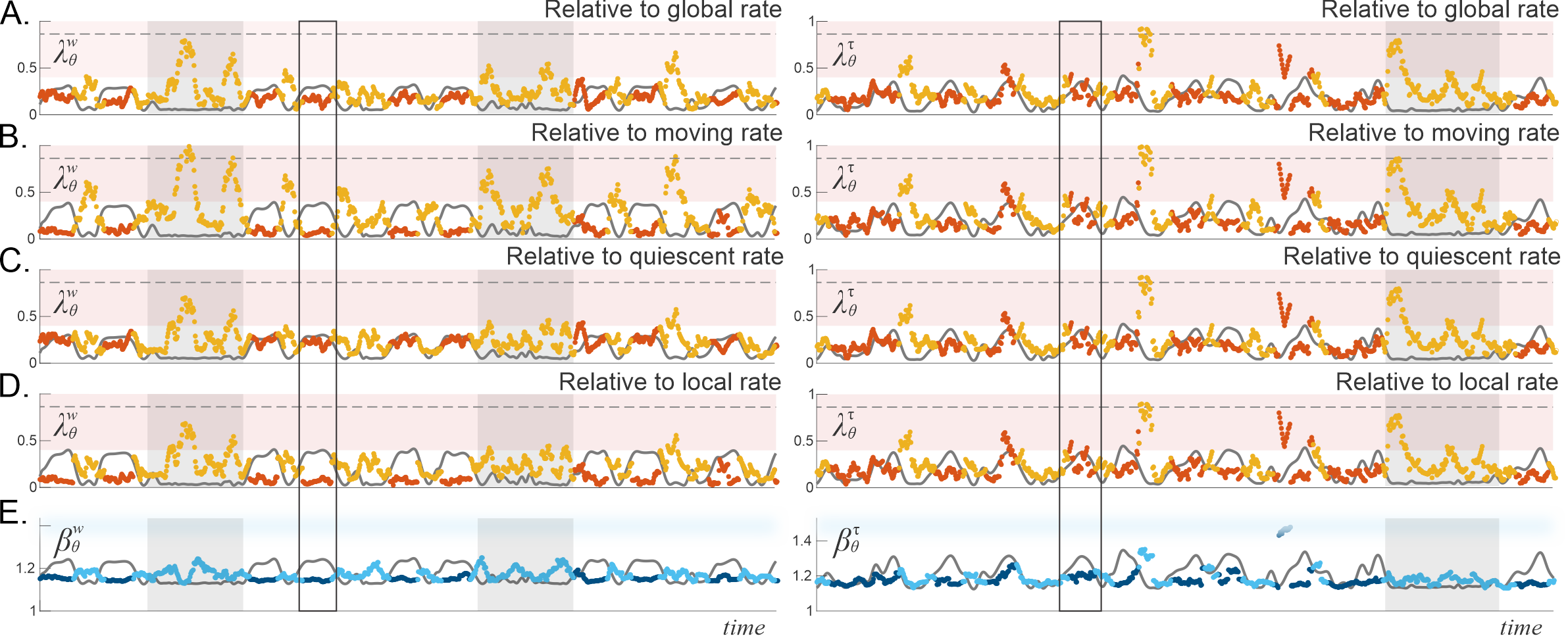}
	\caption{{\footnotesize
			\textbf{Haphazardness of $\theta$-waveforms referenced to different physiological contexts}. The 
			dark red ($\lambda$) and the dark blue ($\beta$) points in all panels represent stochasticity scores
			evaluated during fast-running periods. One such period is highlighted by the vertical black box for
			each phenotype. The yellow ($\lambda$) and the light blue ($\beta$) points correspond to slow
			motions	and quiescence. Prolonged quiescence is emphasized by gray-shaded backgrounds. 
			\textbf{A}. The $\lambda$-score computed relative to the $\theta$-rate averaged over the entire
			running session, closely follows the animal's speed (gray trace) during runs and dissociates from
			it during quiescence in WT mice (left). In tau-mice (right), coupling with speed is weaker.
			During prolonged quiescence, the $\theta$-patterns randomize: $\lambda$-values fall into the
			generic range, $0.4<\lambda<1.8$ (pink stripe same as on Fig.~\ref{fig:stoch}B).
			\textbf{B}. $\lambda^w_\theta$-score assessed relative to the $\theta^w$-rate during active runs
			shows strict adherence with expected trend during speed ups and increases during slowdowns. In
			contrast, $\lambda^{\tau}_\theta$-score for tau-mice appear unaffected by the change in reference.
			\textbf{C}. Relative to quiescence, the adherence of $\lambda^w_\theta(t)$-dynamics to speed is 
			even tighter than relative to global mean (panel A), while $\lambda^{\tau}_\theta$-score remains
			unchanged in both frames of reference. 
			\textbf{D}. If $\theta$-rate is assessed relative to movement during movement and relative to
			quiescence during quiescence, then $\theta^w$-patterns get atypically regular during runs but
			remain as randomized during quiescence as before, which suggests an unbiased, context-free
			stochasticity of the latter. Haphazardness of $\theta^{\tau}$-patterns once again remains 
			indifferent to behavior---note the similarity across the right A-D panels. 
			\textbf{E}. The $\beta^\theta$-scores do not depend on the choice of reference behavior.
	}}
	\label{fig:frame}
	%\end{wrapfigure}
\end{figure}

We noticed two immediate properties in the WT-patterns that are absent in the tau-mice, despite the apparent
similarity of their $\theta$-rhythms: 1) coupling between haphazardness of $\theta^w$-waveforms and the animals'
speed, and 2) changes of $\theta^w$-patterns between activity and quiescence (Fig.~\ref{fig:frame}A,B). During
rapid moves, the stochasticity scores are low, $\lambda_{\theta}^w=0.232\pm 0.001$, $\lambda_{\theta}^{\tau}=
0.238 \pm 0.001$, $p<0.001$, much lower than the Kolmogorov mean ($\lambda^{\ast}\approx0.87$), and $\beta_{
	\theta}^w=1.204\pm 0.001$, $\beta_{\theta}^{\tau}=1.198\pm 0.001$, $p<0.001$, which barely exceeds the 
minimal value (Fig.~\ref{fig:frame}E). The tiny probabilities of such values ($e.g.$, $\Phi(\lambda\leq0.23)
\approx8.11\times10^{-10}$, $\Psi_{16}(\beta\leq1.2)\approx6.74\times10^{-4}$, Fig.~\ref{fig:3D}C) show just
how rare nearly-periodic patterns are among all possible waveforms. During slowdowns and quiescence, patterns
randomize (stochasticity scores grow), as reported in \cite{Hoffman}.
%%%%%%%%%%%%%%%%%%%%%%%%%%%%%%%%%%%%%%%

%%%%%%%%%%%%%%%%%%%%%%%%%%%%%%%%%%
The $\theta$-patterns' haphazardness scores discussed above were evaluated relative to the mean oscillatory
rates observed over the entire running session, $\bar{n}_{\theta}^w\approx8.3$, $\bar{n}_{\theta}^{\tau}\approx
8.1$ peaks per second. On the other hand, it is also well-known that not only the amplitude, but also the mean
rate of $\theta$-rhythmicity changes noticeably between activity and quiescence. Hence, we inquired how would
$\theta$-stochasticity be affected by taking the trending changes into account, and computed two 
$\lambda$-scores: one referenced to the mean rhythm exhibited during active explorations ($\bar{n}_{\theta}^w
\approx8.6$, $\bar{n}_{\theta}^{\tau}\approx8.5$ peaks per second, Fig.~\ref{fig:frame}B), and the other
referenced to the mean oscillatory rate over the quiescent periods ($\bar{n}_{\theta}^w\approx7.9$, $\bar{n}_
{\theta}^{\tau}\approx8.2$ peaks per second, Fig.~\ref{fig:frame}C), for each phenotype. The results demonstrate
that the correspondence between a pattern's ``ongoing haphazardness" and the animals' speed, reported in 
\cite{Hoffman}, is manifested most clearly relative to quiescence. In the ``moving frame of reference," i.e.,
relative to the rate produced during active moves, $\lambda$-scores are low during runs (foreseeably, since 
now these patterns are expected by design), with the mean value $\bar{\lambda}_\theta^w\approx0.125$, and grow
more
%\makebox[\textwidth][s]{now these patterns are expected by design), with the mean value $\bar{\lambda}_\theta^w
%	\approx0.125$, and grow more}
%%%%%%%%%%%%%%%%%%%%%%%%%%%%%%%%%%%%%%%

\begin{figure}[H]
	%	\vspace{-17 mm}
	\centering
	\includegraphics[scale=.84]{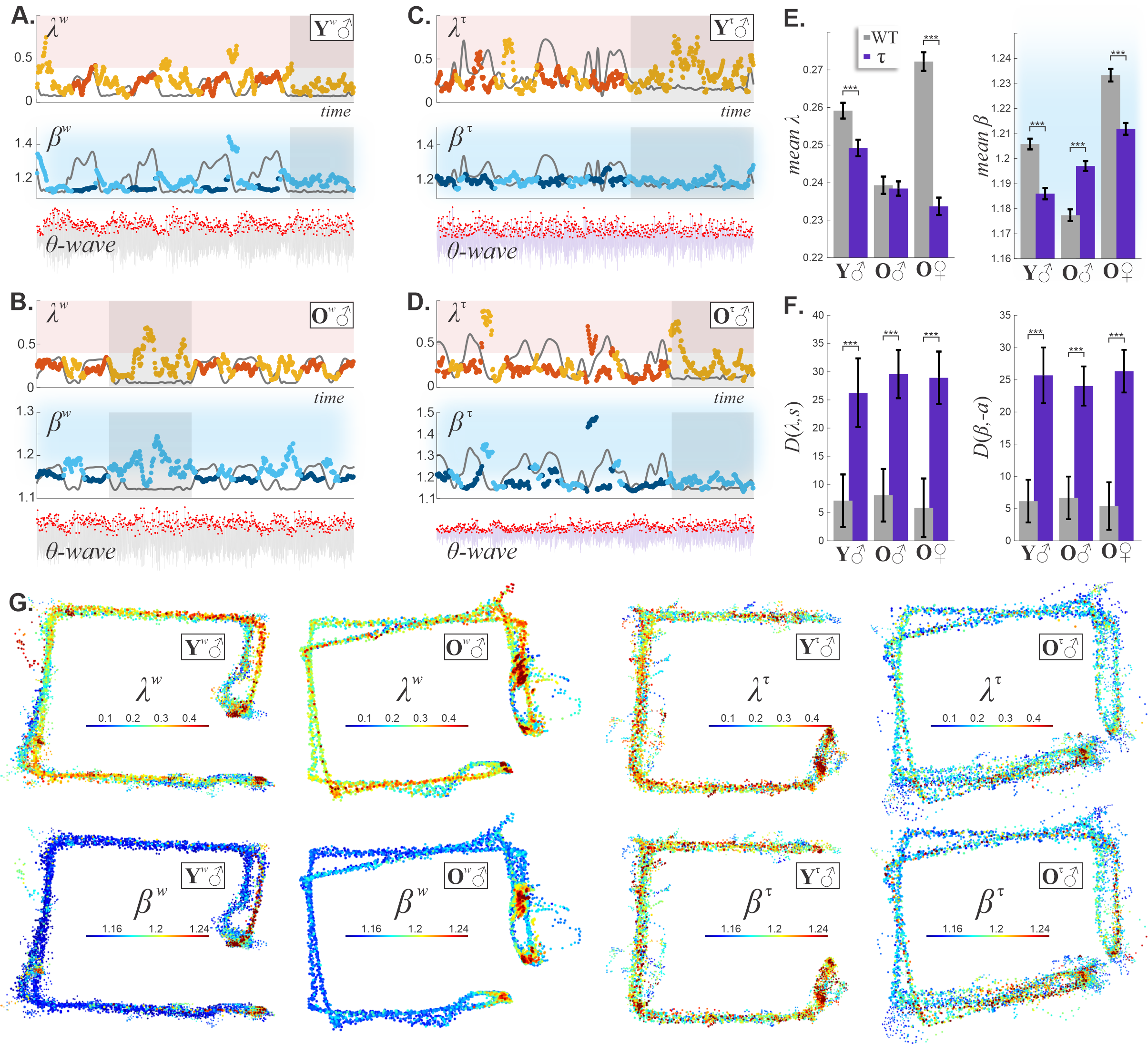}
	\caption{{\footnotesize
			\textbf{$\theta$-patterning and AD progression}.
			\textbf{A}. During lap running, the haphazardness of $\theta^w$-waveforms in young mice shows
			coupling to speed (gray line). Prolonged quiescence (gray bands) diversifies $\theta^w$-patterns.
			The horizontal pink band shows the range of ``typical" $\lambda_\theta^w$-scores, same as on 
			Fig.~\ref{fig:stoch}B. 
			The middle panel shows the dynamics of orderliness of $\theta^w$-patterns. Blue stripe outlines the
			domain of generic $\beta_\theta^w$-values. During fast moves, $\theta$-waves are close to periodic 
			($\beta_\theta^w\approx 1.1$) and slightly more disordered during protracted rest.
			\textbf{B}. In the $9$-month old WT mice, $\theta^w$-waveforms are similar to those in young mice: 
			quasiperiodic (low $\beta_\theta^w$), adherent to the mean (low $\lambda_\theta^w$), coupled to
			speed during lap running, randomizing during quiescence, and disordering during	slowdowns.
			\textbf{C}. In young tau-mice, $\theta^{\tau}$-patterns are more haphazard (higher $\lambda_
			\theta^{\tau}$), but strangely,	less disordered ($\beta_\theta^{\tau}$ bottom panels) and do not
			discriminate between the running and the resting states.
			\textbf{D}. Old tau-mice exhibit stronger discordance between $\theta^{\tau}$-patterning and
			locomotion. 
			\textbf{E}. There is a significant difference between the mean $\lambda$s in $Y_{\male}$ ($n^w=5$,
			$n^{\tau}=4$ mice) and $O_{\female}$ ($n^w=4$, $n^{\tau}=5$ mice) cohorts. Orderliness of 
			$\theta$-patterns differs significantly between all animal groups.  
			\textbf{F}. LCSS distance between $\lambda_\theta$ and speed, as well as between $\beta_\theta$
			and acceleration, is significantly larger in tau-mice. 
			\textbf{G}. Healthy mice demonstrate spatial specificity of $\theta$-patterns, consist between
			laps. In contrast, unsystematic layout of $\theta^{\tau}$-waveforms begins prior to locomotive
			dysfunction in $Y^{\tau}\male$ mice and grows with age (data shown for $n^w=5$, $n^{\tau}=6$ 
			$O^{\tau}_{\male}$ animals). 
	}}
	\label{fig:ThStoch}
\end{figure}

%%%%%%%%%%%%%%%%%%%%%%%%%%%%%%%%%%
\noindent
than twice, to $\bar{\lambda}_\theta^w\approx0.28$, during slowdowns. In other words, relative to quiescence,
the faster the animal moves, the less anticipated, more haphazard $\theta$-waveforms s/he generates, whereas
relative to activity, such waveforms are stipulated---the chance of cycling squarely at the expected mean rate
is vanishing $\Phi(\lambda\leq 0.125)=2.5\times10^{-33}$. Note here that $\beta$-scores do not depend on the
``reference frame," i.e., orderliness is always impartial. 

In contrast, the hippocampi of tau-mice produce much less structured patterns in both reference frames. Despite some
difference in the mean oscillatory rates during active movement and quiescence ($8.2$ vs. $8.5$ peaks per second
respectively, Fig.~\ref{fig:supp2}), the tau-associated $\theta$-rhythm produces same $\lambda^{\tau}$-dynamics,
whether it is referenced to quiescence or to movement. Unlike the speed-coordinated $\theta^w$-patterns,
$\theta^{\tau}$-patterns largely ignore the speed dynamics, regardless of reference trend. Even referencing to
the ongoing, instantaneous rate reveals no behavioral dependence of the $\theta^{\tau}$-stochasticity, which
only increase disorder ($\beta^w\approx1.17$ vs. $\beta^{\tau}\approx1.20$, $p<0.001$).

This stark contrast with the behaviorally contingent $\lambda$-scores of WT mice exposes a curious dissociation
of tauopathic network activity from behavior (Fig.~\ref{fig:frame}B-D, right panels). These effects are 
particularly salient in old animals, which may be due to the net burden on the animal's physical abilities,
as evidenced by lower speed, frequent pausing, etc. (Fig.~\ref{fig:ThStoch}D, \cite{Joshi,Fuhrmann,Bender}). 
However, weakened coupling of $\theta$-waveforms to speed and acceleration are also present in young
tau-mice, who do not exhibit locomotive idiosyncrasies (Fig.~\ref{fig:ThStoch}C,D). In fact, there is a
highly significant drop of the mean stochasticity scores in all tau-mice's (except for the haphazardnesses
in $O_{\male}$ mice)---a feature that clearly separates the two phenotypes (Fig~\ref{fig:ThStoch}E).
%%%%%%%%%%%%%%%%%%%%%%%%%%%%%%%%%%%%%%%

%\begin{figure}[h]
\begin{wrapfigure}{c}{0.5\textwidth}
	\centering
	\includegraphics[scale=0.9]{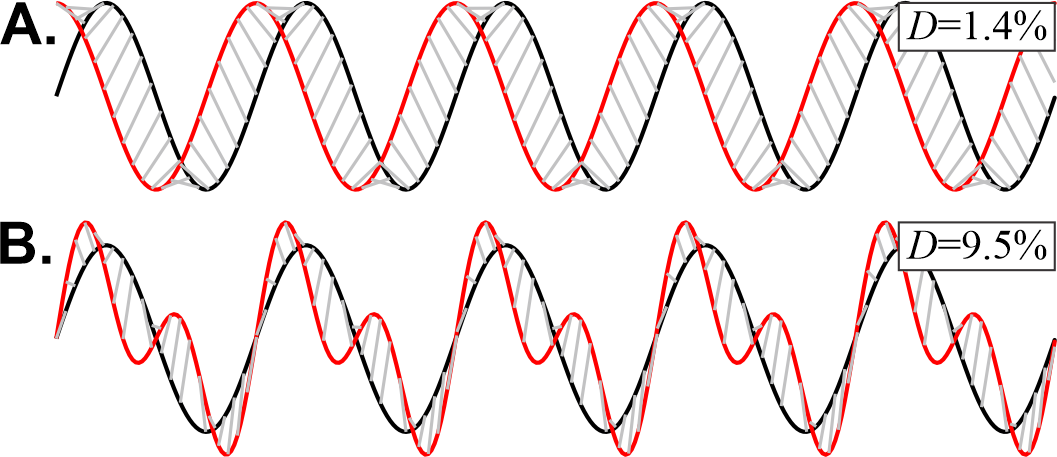}
	\caption{{\footnotesize
			\textbf{Locomotion}. 
			\textbf{A}. LCSS distance between $f_1=\sin(x)$ and $f_2=\cos(x)$ is $D(f_1,f_2)=1.4\%$, which
			reflects the horizontal adjustments required to match the two sequences.
			\textbf{B}. LCSS distance between $f_1=\sin(x)$ and $f_4=\sin(x)+\sin(2x)$ is more than six times
			higher, $D(f_1,f_4)=9.5\%$, due to the increased mismatch in the functions' shapes. Shown is the
			best match achieved after optimal horizontal and vertical scalings and alignments. 
	}}
	\label{fig:lcsstext}
\end{wrapfigure}
%\end{figure}

%%%%%%%%%%%%%%%%%%%%%%%%%%%%%%%%%%
\textbf{Similarity between stochasticity dynamics, speed and acceleration} can be quantified using the Longest
Common Sub-Sequence (LCSS) technique, which allows aligning two profiles through a series of local stretches
\cite{Vlachos,Khan,Morse}. This measure is robust, as it avoids excessively irregular portions of the data 
series, is defined in relative terms (percentage difference between the compared functions) and is intuitive.
For example, there is no LCSS difference between properly aligned sinusoids, $e.g.$, for $f_1=\sin(x)$, $f_2=2
\sin(x)$ and $f_3=\sin(2x)$, one gets $D(f_1,f_2)=0$ and $D(f_1,f_3)=0$, even with some noise added
(Fig.~\ref{fig:lcss}A), while combinations of distinct sinusoids exhibit differences (Fig.~\ref{fig:lcsstext}).
In other words, LCSS accounts for qualitative, essential mismatches between profiles, ignoring trivial shifts,
stretches, and jitters (Fig.~\ref{fig:lcss}B). 

With this in mind, the mean LCSS separation between the $\theta$-haphazardness and speed, $s(t)$, is twice as
large in tau-mice than in WT mice (Fig.~\ref{fig:ThStoch}F). Likewise, the dissimilarity between 
$\theta$-orderliness, $\beta_{\theta}^w(t)$, and acceleration, $a(t)$, is about twice larger in tau-mice.
In comparison, the Fourier power of the $\theta$-band and its frequency drop by $\sim27\%$ and $\sim5\%$ 
respectively ($P^w=0.92$ vs $P^{\tau}=0.67$, $p=0.362$, and $f^w=8.9$ vs $f^{\tau}=8.5$ Hz, $p=0.041$,
Fig.~\ref{fig:spec}). In other words, changes in $\theta$-patterning are much more expressive than the changes
captured by spectral analyses.

\textbf{Spatial maps of stochasticity} are produced by associating the $\lambda(t)$ and $\beta(t)$ scores with
the mouse's ongoing position \cite{Hoffman}. As shown on Fig.~\ref{fig:ThStoch}G, healthy mice, both young and
old, show location-specific $\theta$-patterning, reproduced consistently between laps. Close to the food wells,
$\beta_{\theta}^w$ grows, marking a disordering of the $\theta^w$-waveforms, whereas the quiescence-referenced
$\lambda_{\theta}^w$ decreases, indicating that $\theta^w$-wave recovers its expected behavior as the mouse 
slows down and pauses to eat. Higher $\lambda$-values at the corner nearest to the top food well occur as the
mouse arrives to $F_t$, as opposed to when it departs from it, which marks qualitative changes produced at the
inactivity onset. Likewise, $\beta^w_{\theta}$ maps also show higher values as the mouse transitions from 
movement to quiescence (Fig.~\ref{fig:ThStoch}G). From the quiescence-referenced perspective, patterns randomize
further away from the food wells, in the outer expanses of the track, where $\lambda_{\theta}^w$ increases and
$\beta_{\theta}^w$ drops, indicating the appearance of orderly, but offbeat (relative to reposing expectations)
trains of $\theta$-peaks. In either case, the stochasticity maps of WT mice consistently highlight behaviorally
significant locations along the animal's trajectory.

In contrast, old tau-mice lose these properties altogether: both $\lambda_{\theta}^{\tau}$ and $\beta_{
	\theta}^{\tau}$ become unanchored to the location and scatter arbitrarily along the animal's path 
(Fig.~\ref{fig:ThStoch}H). Even the three-month-old tau-mice, who do not exhibit perceptible changes in
lap running behavior, already have blurred maps of $\theta$-waveforms, which suggests that tau-disturbance
of circuits begins before discernible locomotive abnormalities. These phenomena are even more apparent on the
linearized spatial maps (Fig.~\ref{fig:Laps}), highlighting the failure of tau-mice to follow the spatial
context, which may contribute to spatial cognition and memory dysfunction in the brains of tau mice. In
particular, the coupling between $\lambda^w$-stochasticity and the speed exhibited by WT mice as they 
approach the food
wells is absent in the tau-mice, which may indicate that the former anticipate forthcoming quiescence and
food more than the latter (Fig.~\ref{fig:lamDescent}). 
%\subsection{Spikes}
%\label{sec:spikes}

\textbf{Place cell activity}. The space-specific LFP waveforms in WT mice cannot be explained naively via 
location-specific firing of the hippocampal principal neurons \cite{OKeefe,MosRev}. First, the layout of place
cells in the hippocampus is not topographic, i.e., close principal neurons may respond to far-apart place
fields, and vice versa, cells with proximate place fields may lie far from each other in the network, and
therefore rarely co-contribute to the field detected at a given electrode's tip \cite{Fenton,Redish,Amaral,
	McNaughton}. Second, although the fields' layout---the place field map---controls the order of firing, it
does not define how the spike trains intersperse, i.e., the overall patterning of spikes arriving from different
cells. Furthermore, since place cells' firing is modulated by the mouse's location and speed, i.e., is spatially
and temporally nonuniform, it is unclear which aspect of a given cell's spiking should be attributed to its own
operational stochasticity, and which part is due to the animal's behavioral variability. On the other hand, an
ensemble of cells, whose fields jointly cover the track, provide a roughly continual spiking flow that allows,
$e.g.$, tracking the animal's position in real time \cite{Geisler,Dragoi,Jensen,Brown}. Correspondingly, spike 
patterning in such a flow can be investigated in the same vein as the series of peaks in waveforms.

We studied ensembles of place cells recorded during a single running session, with place field maps
shown on Fig.~\ref{fig:spikes}A, and obtained the spike trains' stochasticity dynamics illustrated on
Fig.~\ref{fig:spikes}B. 
Despite the limited number of simultaneously recorded cells, the scores $\lambda_{P}^w$ and $\beta_{P}^w(t)$
recovered in WT mice trace out a tight motif, pointing at a well-defined, dynamic coupling between spiking 
and locomotion. In other words, spike trains spanning two-second windows ($2-3$ place fields) yield
stochastically consistent spike flow, within the limits of generic haphazardness and the orderliness.

When the animal leaves the top food well (note the rapid approaches to $F_t$ along the vertical protrusion),
$\lambda_{P}^w$-stochasticity increases to very large values, $\lambda_{P}^w\approx10$, which cannot occur by
chance ($\Phi(\lambda\geq10)\approx 0$, Fig.~\ref{fig:spikes}B,C). This implies that the spiking trend changes
qualitatively, as expected: during runs, place cell firing is triggered by the animal's physical traversal
through the \makebox[\textwidth][s]{place fields, whereas in quiescence spiking is produced by ``offline,"
	endogenous network activity}
%%%%%%%%%%%%%%%%%%%%%%%%%%%%%%%%%%%%%%%

\begin{figure}[H]
	\centering
	\includegraphics[scale=.82]{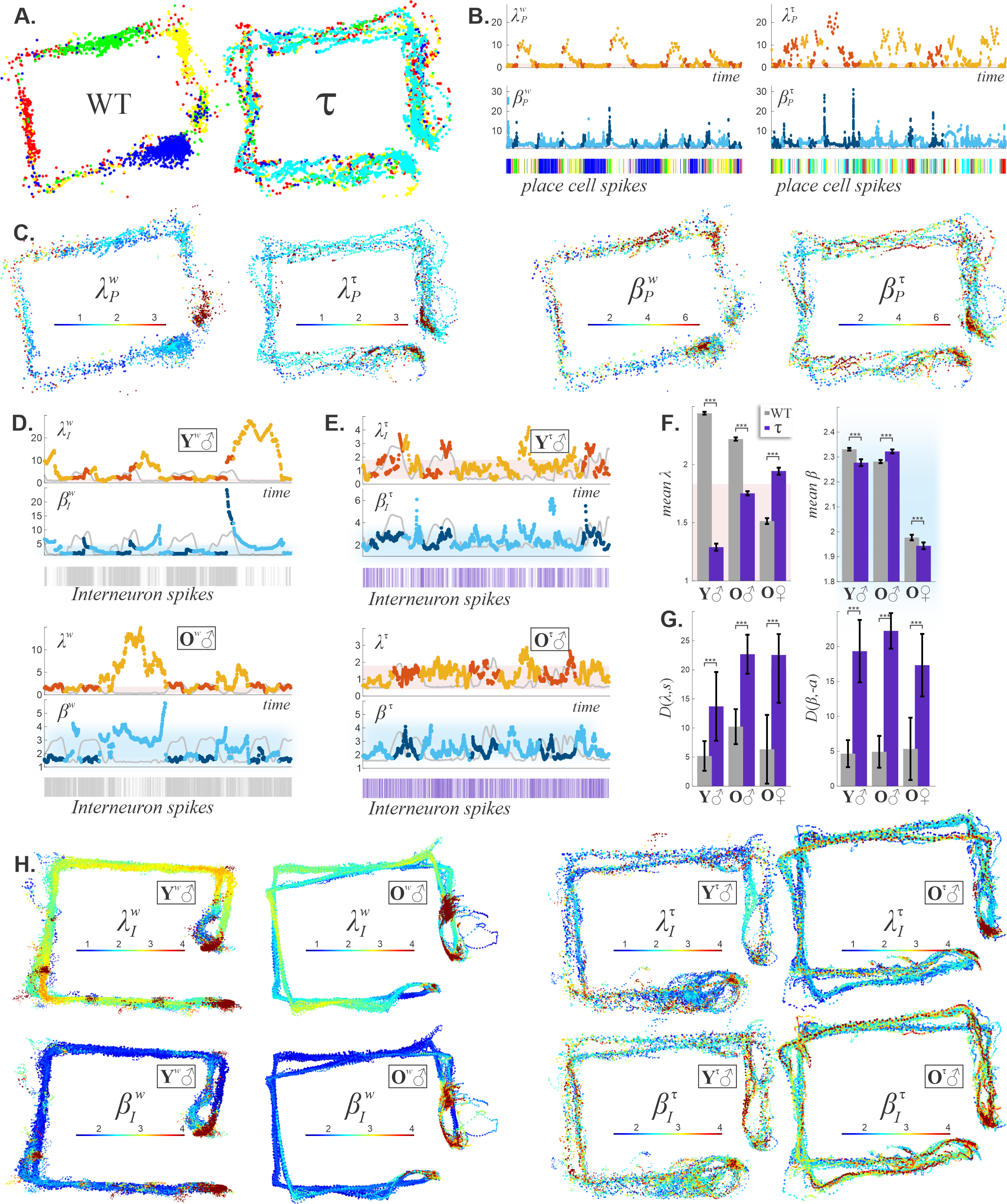}
	\caption{\footnotesize\textbf{Spike stochasticity}. \textbf{A}. Place fields of simultaneously recorded
		$9$-month WT (left) and $9$-month tau-mice (right) place cell ensemble jointly cover the track. Each
		dot represents a spike, color corresponds to a particular place cell, clusters of similarly colored
		dots represent place fields. The WT-fields are clearly defined, the tau-fields are smeared.
		\textbf{B}. In WT, the combined spike flow produced by the ensemble of all recorded place cells forms
		a tight motif.
	}
	\label{fig:spikes}
\end{figure}

%%%%%%%%%%%%%%%%%%%%%%%%%%%%%%%%%%%%%%%
%%%%%%%%%%%%%%%%%%%%%%%%%%%%%%%%%%%%%%%%%
\setcounter{figure}{6}
\begin{figure}[H]
	\centering
	\caption{\footnotesize
		(continued) The values are stochastically generic at the lower food well, $F_b$, and 
		along the track itself. At the other food well, the spike trains become deviant and exhibit high 
		clustering ($\beta_{P}^w\approx 15$), indicating that, at $F_t$, place cells are engaged in offline,
		bristle, autonomous activity. These dependencies are not observed in tau-mice, implying that their
		rapid spiking activity is subdued. Note the greatest upswings of $\beta$ are dark blue,
		i.e., occur as the mouse is actively running. Below spike times are marked by by vertical streaks,
		colored as dots on panel A.
		\textbf{C}. The maps of haphazardness ($\lambda_{P}$, left pair of maps) and orderliness ($\beta_{P}$,
		right pair) for WT and tau-mice reveal a structural deterioration of place cell activity---segments
		of generic and atypical spike patterns intersperse	throughout the trajectory. 
		\textbf{D}. Haphazardness of interneuronal spike trains in fast running $Y^w_{\male}$ mice is coupled
		with speed during runs, and rises conspicuously during quiescence ($\lambda_{I}^w\approx10$), thus
		indicating either an implausible-by-chance deviation of firing from the ongoing trend or a trend change.
		The spike trains (bottom panels) become sparser and cluster more during the rest periods ($\beta_{I}^w
		\approx4$), but then replete and regularize ($\beta_{I}^w\approx1$) during active laps. Older WT mice
		show similar spike patterns, but on a larger scale, from $\lambda_{I}^w\approx1.5$ and $\beta_{I}^w
		\approx1$ during running, to $\lambda_{I}^w\approx10$ and $\beta_{I}^w\approx5$ during quiescence, 
		which suggests that the increase of spiking stochasticity range is likely to be cause by de-tuning.
		\textbf{E}. In tau-mice, interneurons produce generic spike trains at all times---both $\lambda_{I}
		^{\tau}$ and $\beta_{I}^{\tau}$ are in their respective ``typicality bands," i.e., the ordinance of
		tau-spiking is akin to random re-shuffling. 
		\textbf{F}. Interneuron spike trains differ significantly between WT and tau-mice, in all demographic
		groups.
		\textbf{G}. Regardless of age or sex, the haphazardness and orderliness of WT interneurons responded 
		to behavioral context more than tau-interneurons. 
		\textbf{H}. The spatial patterning of interneuron spike trains in WT mice is modulated by its position. 
		\textbf{I}. Tau-interneurons demonstrate a loss of spatially localized spike patterning}.% Continued caption
\end{figure}

\begin{wrapfigure}{h}{0.52\textwidth}
	\centering
%	\vspace{-3mm}
	\includegraphics[scale=0.82]{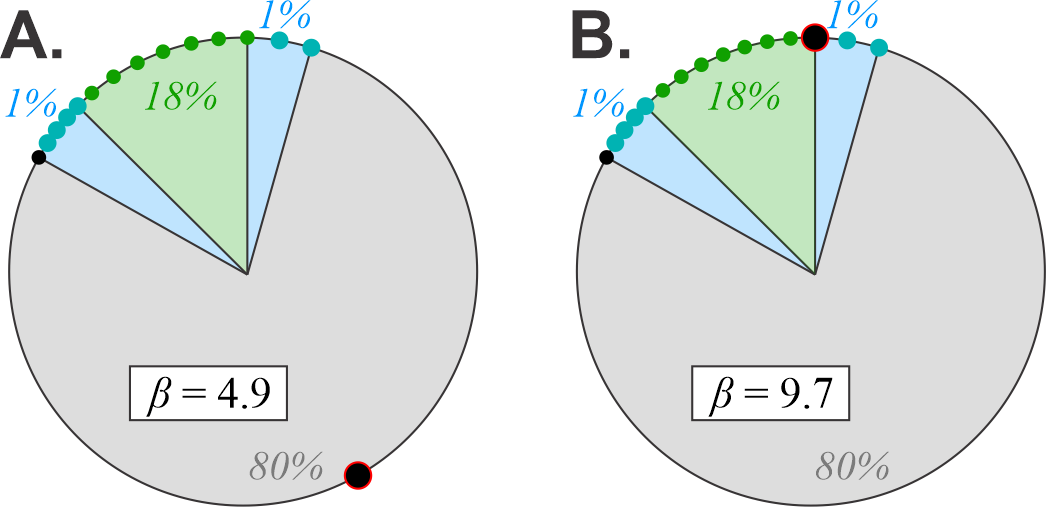}
	%	\vspace{3mm}
	\caption{{\footnotesize
			\textbf{Clustering boosts both $\beta$ and $\lambda$}. 
			\textbf{A}. To achieve high values of $\beta$, the sequence must cluster. As an example, a unit
			circle divided into $4$ segments covering $Q_i={1\%,80\%,1\%,18\%}$ of the circle, and containing,
			respectively, $n_i={2,2,4,7}$ uniformly spaced points that jointly produce $\beta=4.9$. 
			\textbf{B}. By moving a single point from in the sequence from $Q_2$ to $Q_4$ (black dot with 
			red outline),  $\beta$ increases to $9.7$. 
	}}
	\label{fig:clust}
\end{wrapfigure}

%%%%%%%%%%%%%%%%%%%%%%%%%%%%%%%%%%
\noindent\cite{Foster,Karlsson,Carr,Singer,Denovellis}. The latter firings include replays or preplays of
spiking sequences at a rapid timescale, repeated spiking of individual cells, or a combination these activities,
which impact the stochasticity scores \cite{Gupta,Buhry,Stella,Kudrimoti,ONeil,Olafsdottir,Pastalkova,Taxidis}.
Indeed, a replay may last between a couple of dozens to a couple of hundreds of milliseconds, which is 
comparable to the characteristic interval between spikes triggered by the animal's physical passing through
place fields \cite{Karlsson,Nadasdy,Lee,Ji,Diba,Davidson}. These brisk, endogenous network activities clutter
the overall spike patterns and drive the $\beta_P$ and $\lambda_P$ scores up (Figs.~\ref{fig:clust} and 
\ref{fig:burst}). During active movements, the spike disorderliness may also get high, $e.g.$, in the case
illustrated on Fig.~\ref{fig:spikes}B, $\beta_{P}^w\approx 5$ as the mouse approaches food at the top arm,
which also indicates heightened endogenous network activity. 
%%%%%%%%%%%%%%%%%%%%%%%%%%%%%%%%%%%%%%%

%%%%%%%%%%%%%%%%%%%%%%%%%%%%%%%%%%
Tauopathic place cell spiking also changes between generic and cluttering, but it lacks the behavioral 
specificity of WT place cells. In particular, the occurrences of patterns with atypically high $\lambda_{P}
^{\tau}$ and $\beta_{P}^{\tau}$ are not tuned to specific physiological states (Fig.~\ref{fig:spikes}B),
pointing at circuit-level disassociation of collective place cell spiking, both during active behavior and
offline \cite{Ylinen,Csi2,Klaus1}. Thus, not only the individual tau-cells' spatial spiking 
(Fig.~\ref{fig:spikes}A), but also the temporal pattern of the ensemble place cell activity is disrupted
(Fig.~\ref{fig:spikes}B). 

The spiking stochasticity maps also illustrate the disruption of place cell ensemble activity in tauopathy
(Fig.~\ref{fig:spikes}C). In WTs, the distribution of $\lambda^w_P$-values is place-specific, with highs and
lows occurring in the same positions along the mouse's trajectory. Curiously, the WT spike patterns also 
discriminate between the two food wells, exhibiting higher atypicality at one of them. On the other hand, the
place cells of tau-mice produce sporadic, stochastically generic patterns anywhere on the animal's path, and
react to both food wells similarly. 

\textbf{Interneurons} regulate the hippocampal circuit through inhibition, shaping $\theta$-rhythmicity and
other brain waves, in synchrony with the excitatory outputs of principal cells \cite{Squire,Remondes,Manns,
	Moser,Klaus2,Jinno,Varga}. Although the interneurons spike perpetually and at higher frequencies than the
$\theta$-waves oscillate ($\sim20$ Hz vs. $\sim8$ Hz), we found their pattern dynamics to be qualitatively
similar. First, the haphazardness of interneuronal spike trains, $\lambda_{I}^w$, is clearly coupled to speed,
and their orderliness, $\beta_{I}^w$, follows the acceleration (Fig.~\ref{fig:spikes}E). Curiously, spike trains
visibly sparse out during prolonged rest periods, at which times their scores, $\lambda_{I}^w$ and $\beta_{I}^w$,
attain very high, improbable values, indicating trend changes relative to active running (Fig.~\ref{fig:spikes}D).

Also, the pattern range is wider: during active explorations, spike trains are stochastically generic, unlike
the nearly-periodic $\theta$-waves. This suggests that, during lapping, generic interneurons' spiking contributes
to the nearly periodic patterning of $\theta^w$, and during quiescence, clustered spike trains drive spasmodic
$\theta^w$-patterns. Note that, since neuronal spiking is coupled to $\theta$-phase, $\theta$-disturbances may
in turn randomize spiking.

In contrast, neural stochasticity in tau-mice lingers within typical range, both in activity and in 
quiescence (Fig.~\ref{fig:spikes}E). During the former, the stochasticity dynamics alienates from locomotion
($D(\lambda_{I}^{\tau},s)\approx2D(\lambda_{I}^w,s)$; $D(\beta_{I}^{\tau},-a)\approx4D(\beta_{I}^w,-a)$, 
Fig.~\ref{fig:spikes}G). While the WT mice reproduce interneuronal spiking patterns lap after lap, tau-spikes
are largely emancipated from the mice's position, which by itself may be a source of behaviorally incongruent 
$\theta^{\tau}$-patterning (Fig.~\ref{fig:spikes}H). In summary, WT and tau-associated neuronal spikings are
manifestly different: the former are location-specific and well-tuned to speed, while the latter are scattered
all over the track and generic. 

\textbf{$\gamma$-rhythm} ($30-80$ Hz) is the second key LFP component that directs information flow in the
hippocampal network, controls coupling to sensory inputs, synaptic plasticity, mediates attention, and is
involved in a score of other phenomena \cite{Bieri,Jia,Nikoli,ClgMsr,ColginGm}. Previous studies have reported
spectral alterations of the $\gamma$-rhythms in animal models of AD even before plaque formation \cite{Goutagny,
	Iaccarino}, and linked aberrant $\gamma$-activity to cognitive dysfunction \cite{Mably}. 
Indeed, our wavelet scalograms show stronger $\gamma$-power ($40-80$ Hz) in WT mice, particularly during 
periods of active track running between the food wells, while tau-mice do not produce salient $\gamma$
oscillations (Fig.~\ref{fig:spec}). Pattern analyses allow detailing these differences in much more
detail. 

\begin{figure}[H]
	\centering
	\includegraphics[scale=.84]{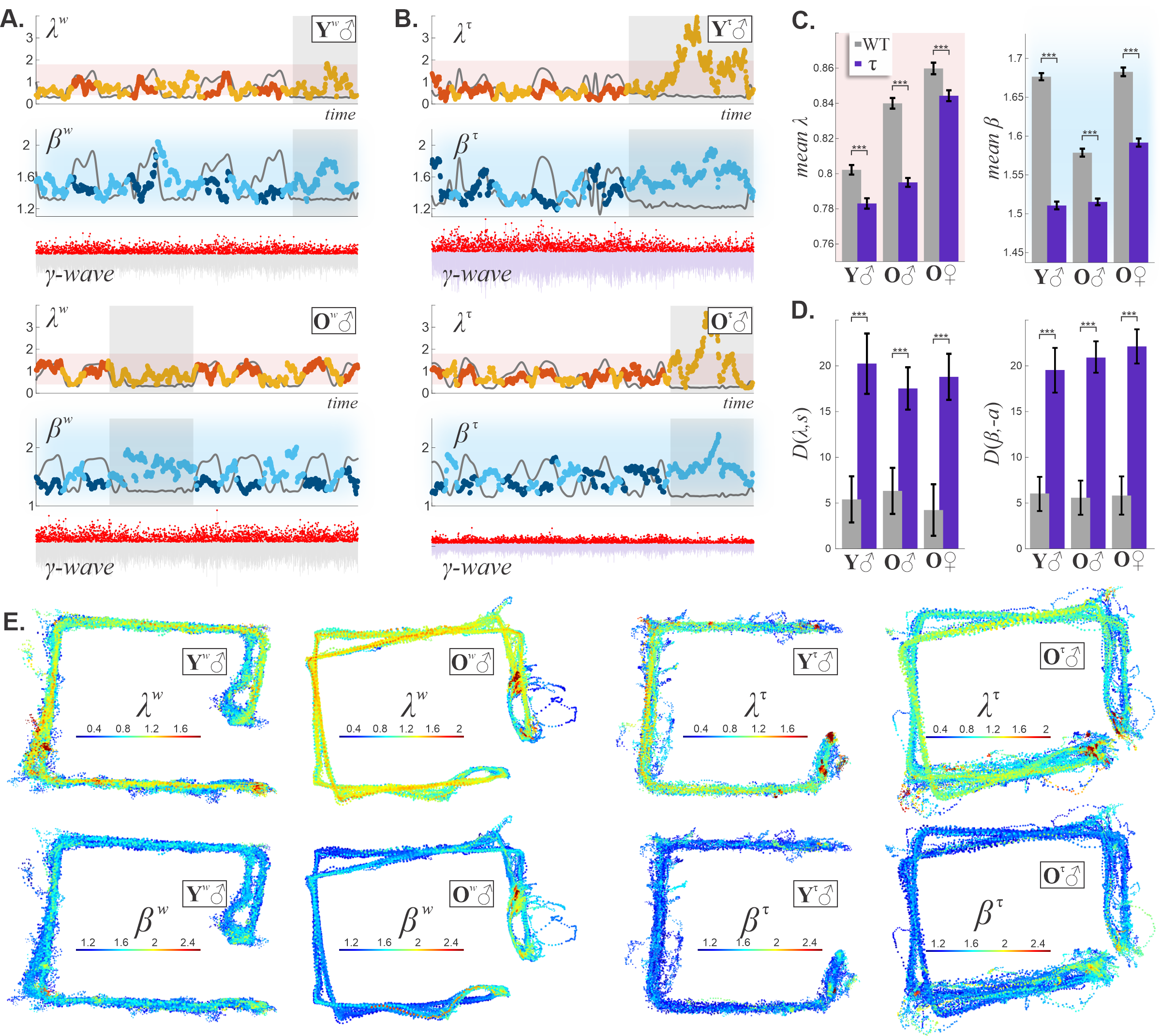}
	\caption{{\footnotesize
			\textbf{$\gamma$-waveform dynamics}. 
			\textbf{A}. In WT mice, $\gamma^w$-waveforms are, for the most part, stochastically generic,
			constrained within the typicality zones, but more diverse than the $\theta$-waveforms (wider
			$\lambda_\gamma$- and $\beta_\gamma$-ranges). During active explorations, $\gamma^w$-patterns
			in $Y^w_{\male}$ mice show coupling to speed (gray trace) and complex dynamics during quiescence
			(gray bands). $\beta_\gamma^w(t)$-scores show that $\gamma^w$-waves are close to periodic during
			activity and disorganize during quiescence, especially at the food wells.
			\textbf{B}. Surprisingly, in tau-mice, the dynamics of $\gamma$-stochasticity is similar: 
			haphazardness of $\gamma^{\tau}$-troughs is also coupled with speed and elevated during quiescence.
			The orderliness, $\beta_\gamma^{\tau}(t)$, likewise, drops during runs and rises to generic levels, 
			$\beta^{\tau}\approx 2$, when s/he stalls.
			\textbf{C}. Mean $\lambda_\gamma$ and $\beta_\gamma$ are significantly lower in tau-mice, in all
			groups---once again, subdued stochasticity is a signature of tauopathy.
			\textbf{D}. The LCSS difference between haphazardness and speed (left) and between orderliness, 
			$\beta_\gamma^w$, and acceleration (right) in WT mice, is more than twice smaller than in 
			tau-mice.
			\textbf{E}. Spatial maps show similar lap-consistent patterning of WT (left panels) and tauopathic
			(right panels) $\gamma$-rhythms.
	}}
	\label{fig:GmStoch}
\end{figure}

To compare $\gamma$-waveforms in WT and tau-mice, we extracted patterns of $\gamma$-troughs that fit into our
selected temporal window, comprised of about $\bar{n}_\gamma^w\approx59$ and $\bar{n}_\gamma^{\tau}\approx57$ 
elements. The first observation, compliant with \cite{Hoffman}, is that $\gamma$-waveforms in WT animals are
much more variegated than $\theta$-waveforms, as indicated by a wider range of the stochasticity scores
(Fig.~\ref{fig:GmStoch}). 
Relative to quiescence, the $\gamma^w$-haphazardness ($\lambda_{\gamma}^w$-score) consistently co-varies with
the speed, while the orderliness, $\beta_{\gamma}^w$, drops to low values during the fast moves. Thus, as with
the $\theta$-waves, $\gamma$-rhythmicity in WT mice has distinct running and quiescent dynamics, with different
appearances in the two frames of reference (Fig.~\ref{fig:GmSlopes}). 

Unexpectedly, $\gamma$-waveforms in tau-mice show similar behavior: the $\lambda_{\gamma}^{\tau}$-scores
also reliably rise and fall with speed, covering about the same range of values, including the periods of 
prolonged quiescence (Fig.~\ref{fig:GmStoch}A,B). However, tau-haphazardness slightly distances from speed,
$D(\lambda_{\gamma}^{\tau},s^{\tau})\approx 3D(\lambda_{\gamma}^w,s^w)$, and the orderliness moves farther
from the acceleration, $D(\beta_{\gamma}^{\tau},-a^{\tau})\approx 3D(\beta_{\gamma}^w,-a^w)$
(Fig.~\ref{fig:GmStoch}D), which may reflect weaker hippocampal coordination of motor effort.
%\makebox[\textwidth][s]{and fall with speed, covering about the same range of values, including the periods of prolonged}

The spatial maps of $\gamma$-stochasticity also maintain consistency between laps, in all mice
\makebox[\textwidth][s]{(Fig.~\ref{fig:GmStoch}E). Near the food wells, $\gamma$-waveforms in both WT and
	tau-mice assume shapes typical for}
%%%%%%%%%%%%%%%%%%%%%%%%%%%%%%%%%%%%%%%%%%%%%%%%%%%%%%%%%%%%%%%%%%%%%%%%%%%

%%%%%%%%%%%%%%%%%%%%%%%%%%%%%%%%%%
\noindent quiescence, despite heightened disorder in this area (smaller $\lambda^{x}_\gamma$, bigger $\beta
^{x}_\gamma$ values). Conversely, series of $\gamma$-troughs consistently become more haphazard and yet more
orderly (take nearly-sinusoidal forms), as the mouse rounds the track's corner furthest from the food.

Relative to the fast-moving mean, the story can be retold as follows: during runs, $\gamma$-waveforms tightly
match with the pseudoperiodic moving trend ($\lambda^{x}_\gamma$ drop), and during extended rest they scramble
($\lambda^{x}_\gamma$ rises, Fig.~\ref{fig:GmSlopes}B). In tau-mice, $\gamma$-waveforms show similar
behavior-specific alterations, albeit more subdued than their WT littermates. The $\lambda$-maps further
emphasize the dichotomy of $\gamma$-patterns between movement and quiescence. 

The fact that $\theta$-waveforms in tau-mice are affected more than $\gamma$-waveforms also implies that 
neuropathology weakens coherence of $\theta$- and $\gamma$-patternings. A number of studies based on correlative
analyses of $\theta$- and $\gamma$-waves' instantaneous amplitudes, frequencies and phases \cite{Canolty,Sirota,
Pastoll}, show that physiologically functional $\theta/\gamma$-comodulation in WT animals \cite{CanoltyLearn,
Benchenane} gets disrupted at early stages of AD \cite{Goutagny,Zhang,Goodman}. Our results show similar effects
at the level of patterns: the LCSS distance between $\theta$- and $\gamma$-haphazardness, $D(\lambda_\theta,
\lambda_\gamma)$, are significantly larger in tauopathic than in WT mice (Fig.~\ref{fig:ThVGm}A). Curiously,
the differences between and speed and $\theta$- and $\gamma$-haphazardness are comparable in WT mice of all
ages and sexes, whereas in the older tau-mice, speed is weaker coupled to $\theta^{\tau}$- than to 
$\gamma^{\tau}$-waveforms, suggesting that $\theta^{\tau}/\gamma^{\tau}$ decoupling is due to deterioration
of the former.

%\subsection{Ripple Events}
%\label{sec:ripples}
\textbf{Ripple events ($R$)} are perturbations of the high-frequency ($150-250$ Hz) extracellular field
oscillations, which reflect the dynamics of autonomous activity in the hippocampal network \cite{BuzSharp,
Csicsvari1,Joo,Leonard}. For example, $R$ are known to co-occur with replays and preplays, which indicates a
connection with memory processing and offline cognitive activity \cite{Fernandez,Girardeau1,Schreiter,
	Girardeau2,Singer,Roux,Sadowski2,Denovellis}. Several mouse models of AD show disruptions of hippocampal
rippling, including increases and decreases of ripple frequency, amplitude, phase alterations, as well as 
disrupted temporal profile, and even disappearance of the Sharp Wave Ripple (SWR) events \cite{Ylinen,Caccavano,Witton,Ji}. 
Yet, the overall structure of the LFP's high frequency domain, as captured by wavelet and Fourier transforms,
are fairly comparable between both phenotypes (Fig.~\ref{fig:spechigh}), whereas pattern-level differences are
distinct.

To evaluate these differences, we identified ``splashes" of the ripple waves' (amplitudes that exceeded two
standard deviations from the mean) and selected series of these events that fall within our selected time 
window\footnote{Note that the $2\sigma$-criterion for selecting $R$ is much lower than the one used in
	SWR detection (typically $5-7\sigma$), which allows identifying $R$ in tau-mice of both age groups.}.
The resulting $R$ patterns contained, on average, a similar number of items in WT and tau-mice ($\bar{n}
_{R}^w\approx14$ and $\bar{n}_{R}^{\tau}\approx 16$ respectively\footnote{Note similarity with the 
$\theta$-peak numbers.}, Fig.\ref{fig:MeanNumPeaks}). 

\begin{figure}[H]
	\centering
	\includegraphics[scale=.84]{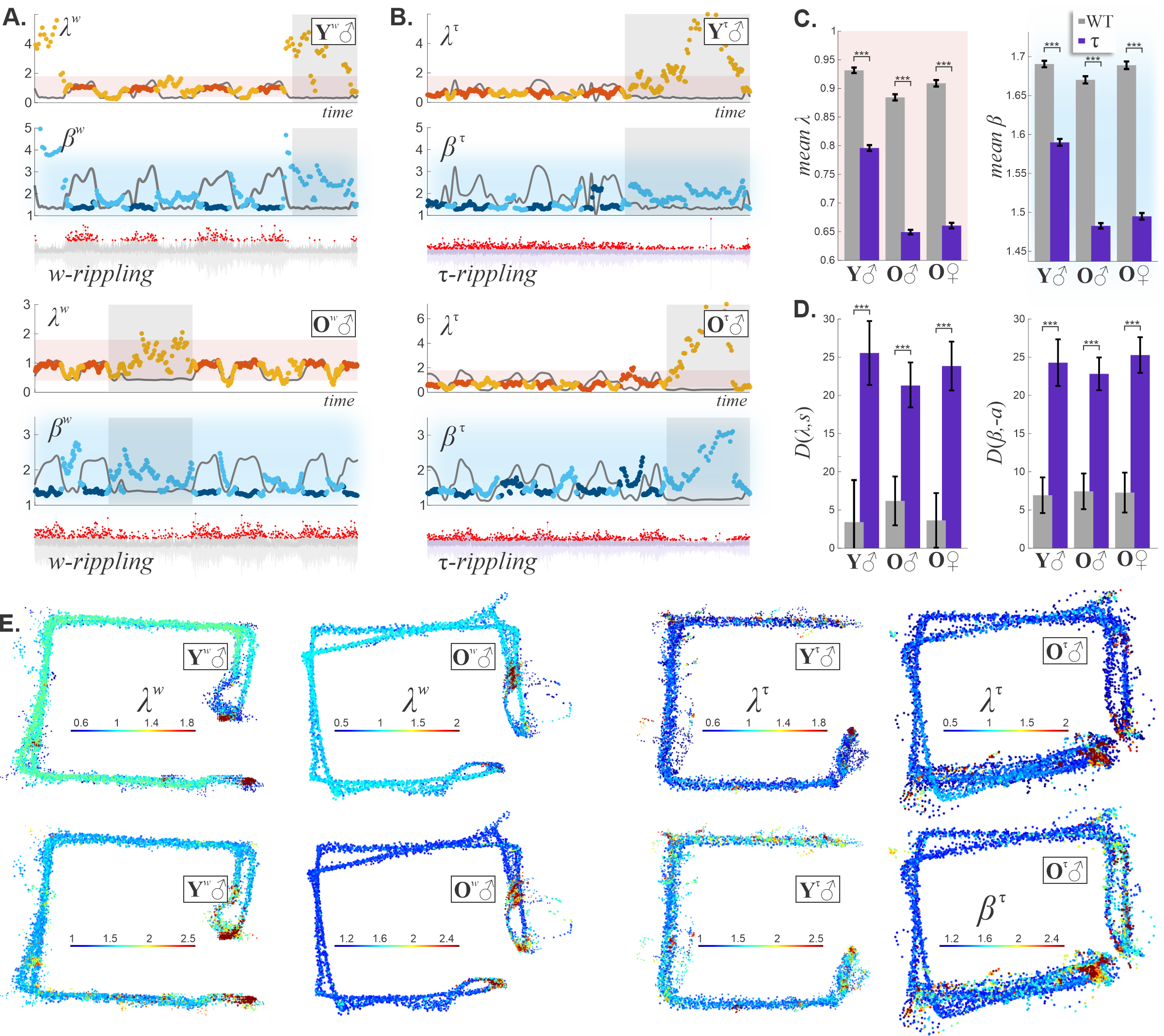}
	\caption{\footnotesize{
			\textbf{Ripple event stochasticity}.
			\textbf{A}. The haphazardness of $R$-patterns in young WT mice (top panel) and old WT mice (bottom 
			panel), is trending during inactivity and scrambles during runs, in a speed-coupled manner. 
			The disorder is also high during quiescence ($\beta_{R}^w\geq 2$), and nearly periodic during movement 
			($\beta_{R}^w\approx1$ at maximal speed).
			(continued) \textbf{B}. In tau-mice, $R$-patterns do not distinguish between moving and quiescent states,
			and are uncoupled from speed. Additionally, both orderliness and haphazardness remain much lower
			than in WTs and do not get atypical, indicating that tau-rippling is stochastically uneventful
			and unresponsive to moves or quiescence. 
			\textbf{C}. WT $R^w$-patterns produce significantly higher $\lambda$- and $\beta$-scores, indicating a 
			much wider range of patterns compared to tau-mice. 
			\textbf{D}. On average, speed is twice closer to $\lambda_{R}^w$ than to $\lambda_{R}^{\tau}$, and
			acceleration is twice closed to $\beta_{R}^w$ that to $\beta_{R}^{\tau}$. 
			\textbf{E}. Spatial maps of $\lambda_{R}^w$ and $\beta_{R}^w$ for both young and old mice reveal 
			spatially specific patterning of $R$. In contrast, spatially scattered $\lambda_{R}^{\tau}$- and
			$\beta_{R}^{\tau}$-values indicate disorder and varied patterns throughout and between laps.
	}}
	\label{fig:RipStoch}
\end{figure}

In accordance with \cite{Hoffman}, the range of WT stochasticity scores referenced to quiescence is wide, 
encompassing both strongly deviating and overly compliant patterns (Fig.~\ref{fig:RipStoch}A). The haphazardness
of rippling in healthy mice, $\lambda_{R}^w$, exhibits clear coupling with speed (Fig~\ref{fig:RipStoch}A,D),
and counter-phase coupling with acceleration. At high speeds, $R^w$ series are nearly periodic, $\beta_{R}^w
\approx1$, and range from nearly-periodic to clustering ($\beta_{R}^w\gtrsim 4$) during quiescence. In contrast,
tau-mice exhibit a smaller assortment of rippling for all age and sex groups ((narrower range of $\beta_{R}
^{\tau}$ and $\lambda_{R}^{\tau}$, opposite to spikes, Fig.~\ref{fig:RipStoch}B,C), and 
more than twice weaker coupling between orderliness and acceleration (Fig.~\ref{fig:RipStoch}D). 
Additionally, $\lambda_{R}^w$ is sensitive to changes of reference trend, while the tauopathic rippling patterns
are indifferent to cross-referencing (Fig.~\ref{fig:RpSlopes}). This loss of coherence between 
$R^{\tau}$-patterning and physiological state, speed or acceleration, suggests that functionality of rapid 
network dynamics is also compromised or lost in tauopathy: tau-rippling is akin to sluggish responding to
random disturbances (generic-lowish $\lambda_{R}^{\tau}$), whereas WT ripples are driven by targeted activity
(Fig.~\ref{fig:clust}).

In terms of spatial maps, both young and old WT mice show clear lap-dependent $R$-patterning as well as a
higher diversity of patterns at the food wells, while tau-mice ripple impartially, with scores deviating
sporadically from the background (Fig.~\ref{fig:RipStoch}E,F, \ref{fig:RpSlopes}).

%%%%%%%%%%%%%%%%%%%%%%%%%%%%%%%%%%%%%%%

%%%%%%%%%%%%%%%%%%%%%%%%%%%%%%%%%%

\newpage
\section{Discussion}
\label{sec:disc}

Most studies focus on cellular AD pathologies or the associated cognitive changes, whereas circuit-level
mechanisms of neurodegeneration receive less attention \cite{DeTure,Perl,Wenk}. This is partly explained
by the immense complexity of the hippocampo-cortical network, intricacy of its dynamics, and multitude of
external inputs, which render detailed, deterministic, causal connections nearly intractable. Furthermore,
understanding the mechanisms by which the individual neurons and synapses contribute to large-scale phenomena,
such as the rhythmicity of extracellular fields, is complicated due to the preponderance of collective,
emergent effects, such as spontaneous synchronization \cite{Arenas1,Liao,Mi,Restrepo,Burton}. Additionally,
concurring neuronal dynamics have different origins and functions. For instance, $\theta$-rhythms are generated 
by medial septal inputs and local interactions between interneurons and principal cells \cite{ColginTh,Bezaire},
whereas $\gamma$-rhythms are produced via local ensemble activity and perisomatic inhibition \cite{Csicsvari1,
	BuzGamma}. 

\begin{figure}[H]
	%\begin{wrapfigure}{R}{0.55\textwidth}
	\centering
	\includegraphics[scale=.75]{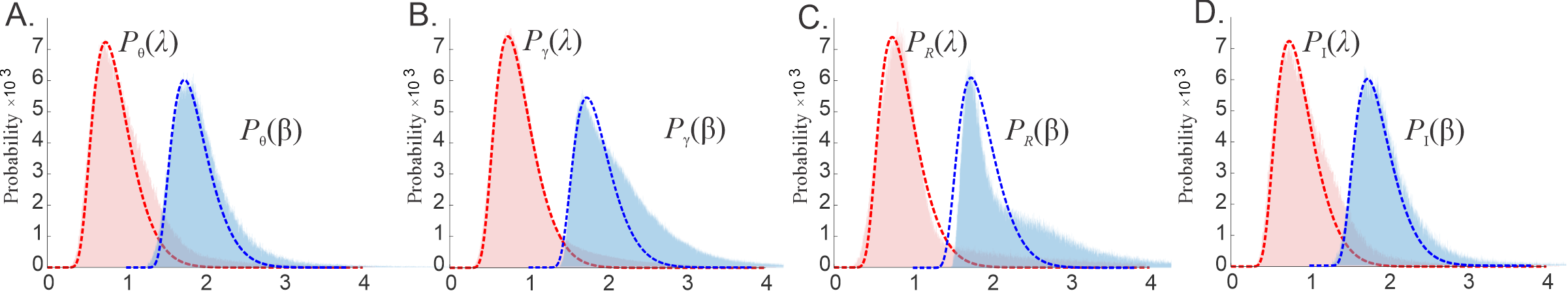}
	\caption{{\footnotesize
			\textbf{Statistical universality of patterns}. Normalized histograms of haphazardness ($\lambda$, pink)
			and	orderliness ($\beta$, blue), evaluated for $1,000,000$ of $25$-point sequences of: 
			\textbf{A}. $\theta$-peak fluctuations;
			\textbf{B}. $\gamma$-peaks;
			\textbf{C}. ripple events;
			\textbf{D}. interneuronal spikes.
			Dashed lines show the shapes of the theoretical Kolmogorov and Arnold distributions, scaled to the
			histograms' heights. These results demonstrate, first, that the varieties of waveforms are broad
			enough to capture the distribution's ranges and shapes in both WT and tau-mice. Moreover, these 
			distributions hold true regardless of data source (i.e. phenotype, age, LFP, single neuron, etc.). 
			In the case of $\gamma$-patterns and ripples,
			$\beta_\gamma$ and $\beta_{R}$ histograms deviate from the impartial $\beta$-statistics, due to 
			larger proportions of cluttering waveforms. Data collected from $14$ WT and	$16$ tau-mice,
			the spikes are recorded from $12$ WT and $5$ tau-interneurons.
			% firing at \textcolor{red}{XX Hz, over $XX$ minute sessions}.
	}}
	\label{fig:univ}
	%\end{wrapfigure}
\end{figure}

Remarkably, despite the differences in spatiotemporal scales, mechanisms and implementations, patterns of neural
activity follow the same universal statistics described by Kolmogorov and Arnold distributions. In other words,
the mathematical laws governing the probability of patterns' appearance overrides the exhaustive physiological
details, much like the sampling distribution of the means is always Gaussian, irrespective of the mechanisms 
that contribute individual inputs (Fig.~\ref{fig:univ}). Consequently, time-variations of stochasticity scores,
$\lambda(t)$ and $\beta(t)$, evaluated for continuously evolving waveforms and spike trains provide an impartial 
description of the circuit dynamics, emancipated from the minutiae.
%%%%%%%%%%%%%%%%%%%%%%%%%%%%%%%%%%%%%%%

%%%%%%%%%%%%%%%%%%%%%%%%%%%%%%%%%%
%\newpage
An important advantage of the $\lambda/\beta$ stochasticity scoring is its intuitive transparency, that enables
reasoning about neural activity in colloquial terms, focusing on structural regularities or peculiarities, $e.g.$,
atypical periodicity, excessive clustering, etc. Due to their universality, these quantities may potentially be
actualized, i.e., represented explicitly in the brain, and used to guide system-level information exchange, 
movement coordination, memory encoding, retrieval and other phenomena. 
%Brain waves modulate neuronal activity and hence contribute to cognition and behavior \cite{Nyhus,Heitmann,
%Rubino,Baker,Sanes}.

Returning to the phenotype comparisons, generic differences between tau pathology and WT cases can be summarized
as follows. While healthy mice exhibit well-tuned, purposeful, behavior-modulated LFP rhythmicity and spiking
patterns, tau-mice show dissociation of neural activity from physiological and behavioral states. The way in 
which the hippocampal spiking is produced, modulated by the brain waves and imparted onto downstream networks is
qualitatively altered by tau-pathology.

These observations are in line with previous studies demonstrating that compromised coupling between LFP rhythms
and locomotion, loss of spatial and temporal fidelity of hippocampal neurons are key contributors to learning
and memory impairments \cite{Jun,Cacucci2,Zhang,Goodman}. We demonstrate that not only do tau-pathologies
disturb individual neurons and rhythms, but they also scramble the overall spatiotemporal architecture of the 
hippocampal activity, which is saliently manifested. The morphological alterations illustrated on 
Fig.~\ref{fig:waves} may indeed be indicative of systemic alterations in circuit activity and can  hence serve
as signatures of neurodegeneration.

\section{Acknowledgments}
	C.H. and Y.D. are supported by NIH grant R01AG074226 and R01NS110806, and by NSF grant 1901338.
	C.J. and D.J. are supported by NIH grants R01MH112523 and R01NS097764.

\newpage

\section{Mathematical Supplement}
\label{sec:math}

\textbf{Computational algorithms} are based on the works \cite{Kolm,Arn1,Arn2,Arn3,ArnB2,ArnB4,Stephens},
and outlined in \cite{Hoffman}.

%\begin{enumerate}[leftmargin=0.25cm]
%\item 
\textit{1. Kolmogorov score}. Let $N(X,L)$ be the \textit{empirical distribution} of an ordered sequence $X=\{
x_1,x_2\ldots x_n\}$---the number of $x$-elements in the interval between $0$ and $L$,
\begin{equation*}
	N(X,L) = \{\textrm{number of} \,\, 0\leq x_k < L\}.
	\label{N}
\end{equation*}
For semiperiodic series, $e.g.$, brain rhythms and regularly appearing spikes, this function grows
proportionally to $L$, with the mean slope defined by the average frequency, $\bar{N}(X,L)=\bar{f}L+m$. The
deviations of a pattern $X$ from this mean, normalized by a ``random walk" magnitude,
\begin{equation}
	\lambda(X)= \sup_L|N(X,L) - \bar{N}(X,L)|/\sqrt{n},
	\label{lam}
\end{equation}
is universally distributed. The probability distribution
\begin{equation}
	P(\lambda) = 2\lambda \sum_{k=-\infty}^{\infty}(-1)^{k+1} k^2 e^{-2k^2 \lambda^2},
	\label{Plam}
\end{equation}
is concentrated between $\lambda^{-}=0.4$ and $\lambda^{+}=1.8$, with the mean $\lambda^{\ast}\approx0.87$
(Fig.~\ref{fig:stoch}B). Outside of these limits, $\lambda$-scores appear with probability less than $0.3\%$,
marking the``atypical" patterns \cite{Kolm,Stephens,Arn1,Arn2,Arn3}.
%This is akin to the familiar use of normally distributed $z$-scores: ``typical" $z$s ($99.7\%$ of them) lay in
%the range $-3\leq z\leq3$, while small ($z<-3$) or large ($z>3$) scores appear with probability $0.3\%$. Just
%as the $z$-scores are used to detect outliers in a normal distribution, $\lambda$-scores measure typicality of
%patterns: if a pattern $X$ deviates from the expected mean within limits established by the ``hump" of the 
%$P(\lambda)$-distribution (pink stripe on Fig.~\ref{fig:stochs}B), then $X$ is typical. If, however, these 
%limits are surpassed, then the pattern is ``atypical," ``uncommon," ``not chancy," and may be viewed as a sign
%of a severe trend violation, or a ``trend change."
Including higher-order corrections to the Kolmogorov score,
\begin{equation}
	\lambda(X)\to\lambda(X)\left(1+\frac{1}{4n}\right)+\frac{1}{6n}-\frac{1}{4n^{3/2}},
	\label{lamn}
\end{equation}
increases statistical accuracy for short sequences ($10-20$ elements) to over $0.01\%$ \cite{Bol1,Vrbik1}.

%\item 
\textit{2. Arnold score}. Arranging the elements of $X$ on a circle of length $L$ yields $n$ consecutive arcs,
with lengths $l_1,l_2,\ldots,l_{n}$ (Fig.~\ref{fig:stoch}C). The combination 
\begin{equation}
	\beta =\frac{n}{L^2}(l_1^2+l_2^2+\ldots+l_n^2)
	\label{B}
\end{equation} 
quantifies the orderliness of $X$: if the points $x_k$ ``repel" each other and strive to maximize separation,
then $\beta$ is small (minimum $\beta=1$ is achieved by perfectly symmetrical, equispaced layouts). If the elements attract
and tend to cluster, then $\beta$ is large (up to $\beta=n$, reached when all $x_k$s collapse into one point).
Randomly placed, independent elements produce $\beta$-scores close to the universal value, $\beta^{\ast}\approx 2$
\cite{ArnB2,ArnB4}.
The length $L$ of the circle was selected so that the distance between the end points, $x_0$ and $x_n$, on the
circle became equal to the mean of the remaining arc lengths, $l_n=|x_n-x_0|_{\mod L}=\bar{l}_i$.

\textit{3. Probability $\beta$-distributions} are parameterized by the number of elements in the sequence. As 
shown on Fig.~\ref{fig:3D}, these distributions have the mean of $\beta_{n}^{\ast}\approx 2n/(n+1)$ and a 
close-by peak, and rapidly decay as $\beta$ approaches $\beta_{\min}=1$ or grows over $\beta^{+}=3.5$. For
intermediate sized samples used in practice, $e.g.$, for $n=50$, $\beta^{-}=1.5$ and $\beta^{+}=3.2$
(Fig.~\ref{fig:3D}D). 
%%%%%%%%%%%%%%%%%%%%%%%%%%%%%%%%%%%%%%%

\setcounter{figure}{0}
\renewcommand{\thefigure}{M\arabic{figure}}
\renewcommand{\theHfigure}{M\arabic{figure}}
\begin{figure}[H]
	%\begin{wrapfigure}{c}{0.7\textwidth}
	\centering 
	\includegraphics[scale=1.2]{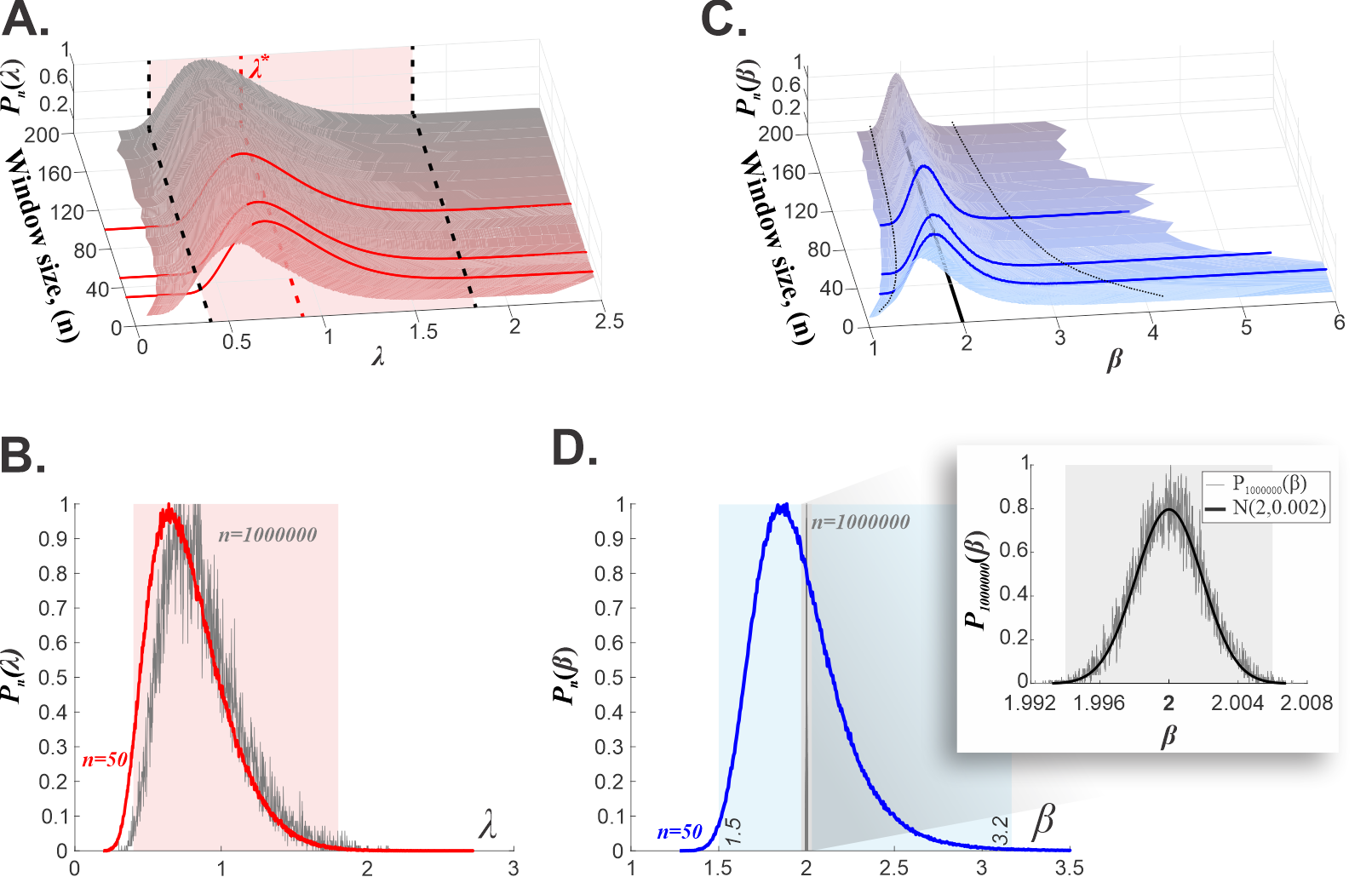}
	\caption{{\footnotesize
			\textbf{Stochasticity scores and sample sizes}. 
			\textbf{A}. The shape of the $\lambda$-histogram depends weakly on the sample size. If the latter 
			exceeds $n\approx 25$, the histogram closely matches the theoretical shape (\ref{Plam}). The shadowed
			domain contained between dashed lines contains $99.7\%$ of values.
			\textbf{B}. Empirical $\lambda$ distributions for a $50$-point sample (red) compared to a $1000000$-point
			sample (gray) reveals little difference. 
			\textbf{C}. $\beta$-distributions narrow as the number of samples per window increases. The dashed
			lines in the horizontal plane ($n-\beta$ axes) mark the upper and the lower bounds, $\beta_{n}^{+}$
			and	$\beta_{n}^{-}$, which contain $99.7\%$ of $\beta$-values for each sample size. Shown are the
			$\beta_n$-histograms numerically evaluated for $20\leq n\leq 200$. In terms of the relative size,
			$x=n/200$, the polynomial fits for the limiting curves produce $\beta_{x}^{+}\approx 4.5-8.4x+16x^2
			-15.2x^3+5.44x^4+O(x^5)$, with $R^2=0.9994$ and	$\beta_{x}^{-}\approx 1.1+2.2x-4.0x^2+3.92x^3-1.42
			x^4+O(x^5)$ with $R^2=0.9995$. 
			\textbf{D}. Comparing $\beta$ distributions for a $50$-point sample with typical range (shadowed 
			blue) to the empirical distribution of $1000000$-sample sequences (gray) reveals significant
			narrowing. The pop-out represents the zoomed-in very narrow $1000000$-point $\beta$-distribution,
			approximated by a Gaussian distribution with mean $2$ and standard deviation $0.002$ (black line).
	}}
	\label{fig:3D}
	%\end{wrapfigure}
\end{figure}

\begin{figure}[H]
	%\begin{wrapfigure}{c}{0.55\textwidth}
	\centering
	\includegraphics[scale=1.25]{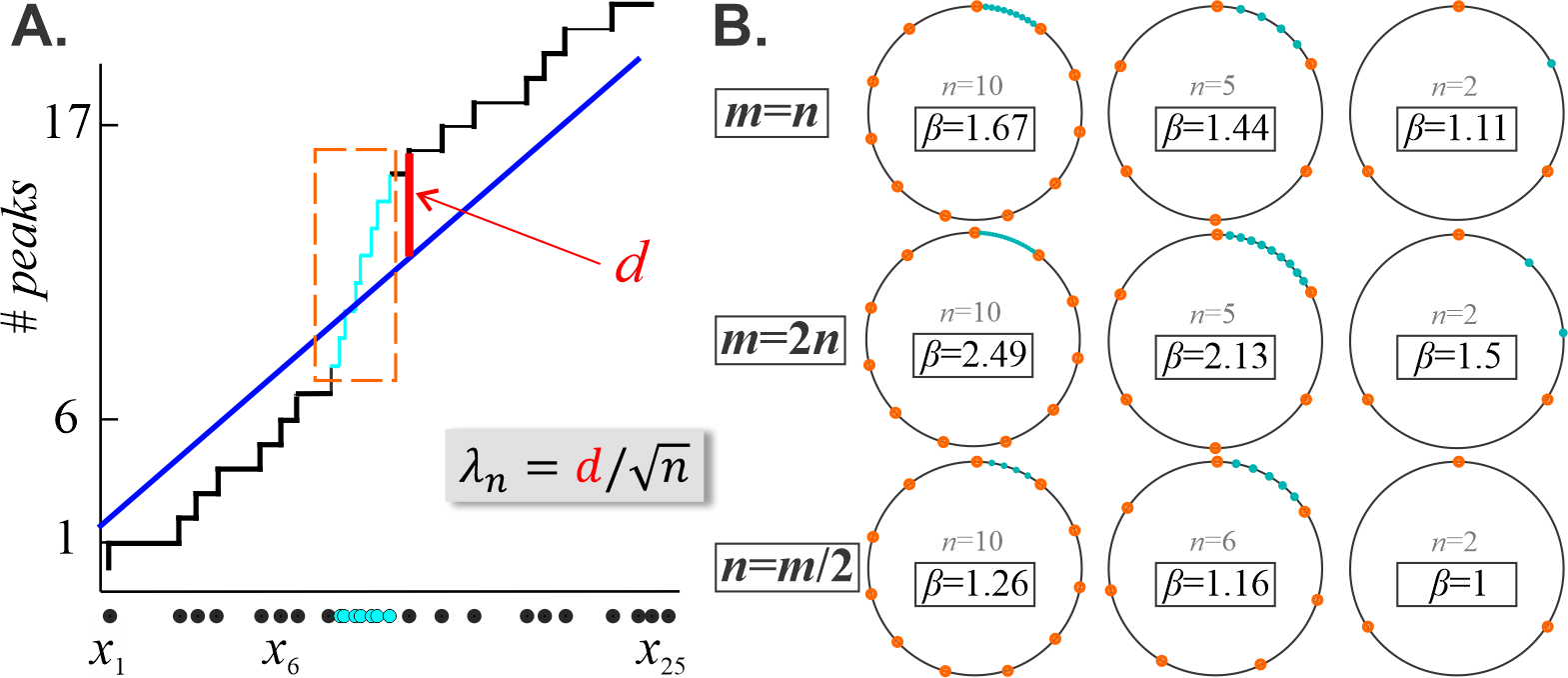}
	\caption{{\footnotesize
			\textbf{Clustering segments increase stochasticity scores}.
			\textbf{A}. The empirical distribution built for a sequence of $17$ points (Fig.~\ref{fig:stoch}A),
			with additional seven points inserted (blue segment of the staircase). The latter raises the
			intercept without altering significantly the slope of the mean (blue line), which results in a
			jump-up of haphazardness score, $\lambda$.
			\textbf{B}. The circle is partitioned into $n+1$ equal arcs, with one further split into $m$ segments. 
			Varying the values of $n$ and $m$, or adjusting the proportion of each, produces ``clustering" that 
			affects the value of $\beta$.
		}
	}
	\label{fig:burst}
	%\end{wrapfigure}
\end{figure}

%%%%%%%%%%%%%%%%%%%%%%%%%%%%%%%%%%%
\textit{4. Rapid replays increase $\lambda$ and $\beta$}. A dense sequence of points creates a ``bulge" in
the empirical distribution, hence resulting in a $\lambda$ that deviates farther from the expected mean
(Fig.~\ref{fig:burst}A). Similarly for the orderliness: if $n$ points are distributed uniformly on the circle,
$l_k=L/n$, $k=1,2,\ldots,n$, then $\beta=1$. If one arc is additionally split by $m-1$ extra points into $m$
smaller equal arcs of lengths $\tilde{l}_{i}=L/nm$, $i=1,2,\ldots,m$ then
\begin{equation*}
	L =\sum_{k=1}^{n-1}l_k +\sum_{i=1}^{m}\tilde{l}_{i},
	%\label{Ba}
\end{equation*}  
and (\ref{B}) produces
\begin{equation}
	\beta =\frac{(n-1)+m}{L^2}\left(\sum_{k=1}^{n-1} l_{k}^2+\sum_{i=1}^{m}\tilde{l}_{i}^2\right)
	=\frac{n+m-1}{n}\left(\frac{n-1}{n}+\frac{1}{nm}\right).
	\label{Ba}
\end{equation} 
As an illustration, if $m=1$ (no points added), then $\beta=1$. If $m=n+1$, then (\ref{Ba}) reduces to 
\begin{equation}
	\beta =2\frac{n}{n+1},
	\label{b1}
\end{equation}
which, coincidentally, matches the ``impartial mean" $\beta^{\ast}$ \cite{ArnB2,ArnB4}. If $m=2n+1$, then
formula (\ref{Ba}) yields
\begin{equation}
	\beta=3\left(1-\frac{2}{2n+1}\right),
	\label{b2}
\end{equation} 
and if $m=n/2+1$ (assuming $n$ is even), then
\begin{equation*}
	\beta %=\frac{3}{2}\left(\frac{n^2+2}{(n+1)^2}\right)
	=\frac{3}{2}\left(1-\frac{1}{n+2}\right),
	%	\label{Ba}
\end{equation*} 
etc. In general, if $p$ arcs, $l_{k_1},l_{k_2},\ldots,l_{k_p}$, are subdivided by additional $m_1-1$, $m_2-1$,
..., $m_p-1$ points, uniformly laid out, they yield, respectively, $m_1$ sub-arcs of length $\tilde{l}_{k_1}=
L/(nm_1)$, $m_2$ sub-arcs of length $\tilde{l}_{k_2}=L/(nm_2)$ and so forth. Correspondingly, 
\begin{eqnarray}
	\beta &=&\frac{(n-p)+m_1+\ldots+m_p}{L^2}\left(\sum_{k\neq k_j}l_{k}^2+\sum_{i_1=1}^{m_1}\tilde{l}_{i_1}^2
	+\ldots+\sum_{i_p=1}^{m_p} \tilde{l}_{i_p}^2\right)\nonumber\\
	&=&\frac{n+m_1+\ldots+m_p-p}{n^2}\left(n-p+\sum_{j=1}^{p}\frac{1}{m_j}\right). 
	\label{Bp}
\end{eqnarray}
By varying proportions between $m_j$ and $n$, changing the number of divided segments, etc., one can produce
different values of $\beta$ to model clustering densities discussed in Sec.~\ref{sec:res}, as illustrated below.
\vspace{10pt}
\begin{enumerate}[nosep,leftmargin=1.7cm]
	\item[\textbf{Case 1.}]\label{c1} If the number of extra points on $l_{k_j}$ is $m_j=n+1$, then
	(\ref{Bp}) reduces to
	\begin{equation}
		\beta =(p+1)\left(1-\frac{p}{n(n+1)}\right),
		\label{bb1}
	\end{equation}
	i.e., each split segment contributes $\Delta\beta\approx1$ to raising the overall $\beta$, as implied by
	(\ref{b1}). 
	\item[\textbf{Case 2.}]\label{c2} If there are $m_j=2n+1$ points in each divided segment, then (\ref{Bp}) 
	yields
	\begin{equation}
		\beta=(2p+1)\left(1-\frac{p}{n(2n+1)}\right),
		\label{bb2}
	\end{equation} 
	i.e., the effect described by (\ref{b2}) is produced $p$ times.
\end{enumerate}
\vspace{10pt}
If the sliding window contains $N$ data points total,
\begin{equation*}
	N=n-p+\sum_{j=1}^p m_j,
\end{equation*}
then one gets $N=n(p+1)$ in Case 1 and $N=n(2p+1)$ in Case 2. Specifically, if $N=15$ as with the $\theta$-peaks
or ripple events, then $15=n(p+1)$ and $15=n(2p+1)$. With two inserted subsequences, $p=2$, the first equation 
necessitates $n=5$ regular points, and the second $n=3$ points, yielding $\beta=2.85$ (Eq.~\ref{bb1}) and $\beta
=4.5$ (Eq.~\ref{bb2}).
%%%%%%%%%%%%%%%%%%%%%%%%%%%%%%%%%%%%%%%

%%%%%%%%%%%%%%%%%%%%%%%%%%%%%%%%%%

\textit{4. Longest common subsequence}. Comparing shapes is a challenging problem. Computing point-by-point 
Euclidean distances may produce misleading results, especially when signals contain noise or if there are 
misalignments between pairs of points. More intuitive results are produced by nonlinear alignments that allow 
for elastic shifting of points without rearrangement, $e.g.$ longest common subsequence (LCSS) \cite{Vlachos,Khan,Morse}.

%%%%%%%%%%%%%%%%%%%%%%%%%%%%%%%%%%%%%%%

\begin{figure}[H]
	\centering
	\includegraphics[scale=1]{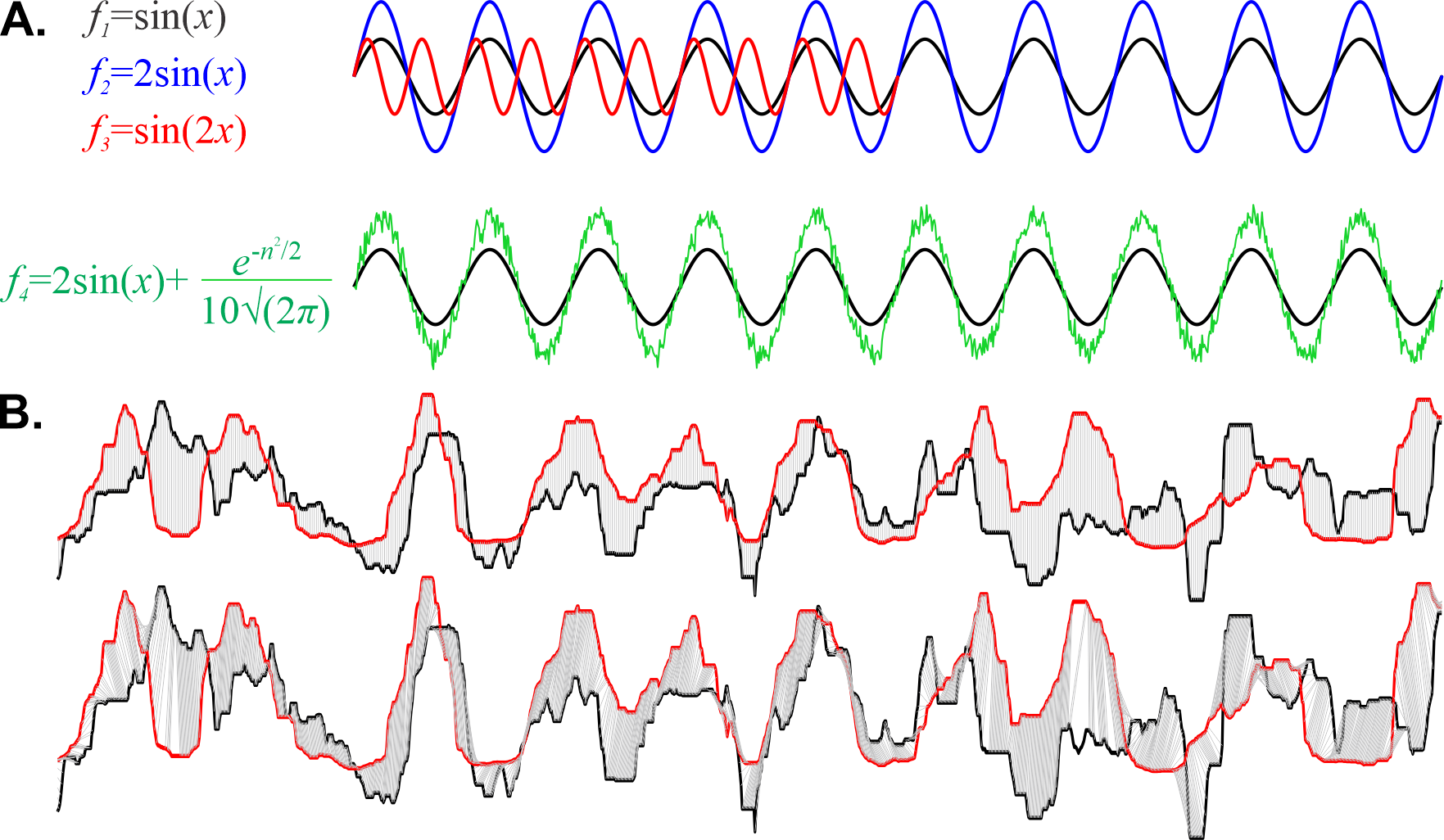}
	\caption{{\footnotesize
			\textbf{LCSS example}. 
			\textbf{A}. LCSS discounts shifts or amplitude changes (top), as well as added noise---in all
			illustrated cases the LCSS distance vanishes.
			\textbf{B}. Euclidean distance for vertical point-to-point matching on example $\lambda$ vs speed 
			data can result in poor shape comparison. 
			\textbf{C}. Oblique LCSS matching (top) allows shifting of the time axis for more intuitive comparisons 
			of the speed vs $\lambda$ profiles. The bottom panel shows the LCSS matched shapes.
			%This technique does not force shape matching by requiring every point to be matched. 
	}}
	\label{fig:lcss}
\end{figure}

\newpage
\section{Supplementary Figures}
\label{sec:sfigs}

\setcounter{figure}{0}
\renewcommand{\thefigure}{S\arabic{figure}}
\renewcommand{\theHfigure}{S\arabic{figure}}

%%%%%%%%%%%%%%%%%%%%%%%%%%%%%%%%%%%%%%%
% SUPPLEMENTARY FIGURE 1
\renewcommand{\figurename}{Fig}
\begin{figure}[H]
	%\begin{wrapfigure}{c}{0.7\textwidth}
	\centering
	\includegraphics[scale=.8]{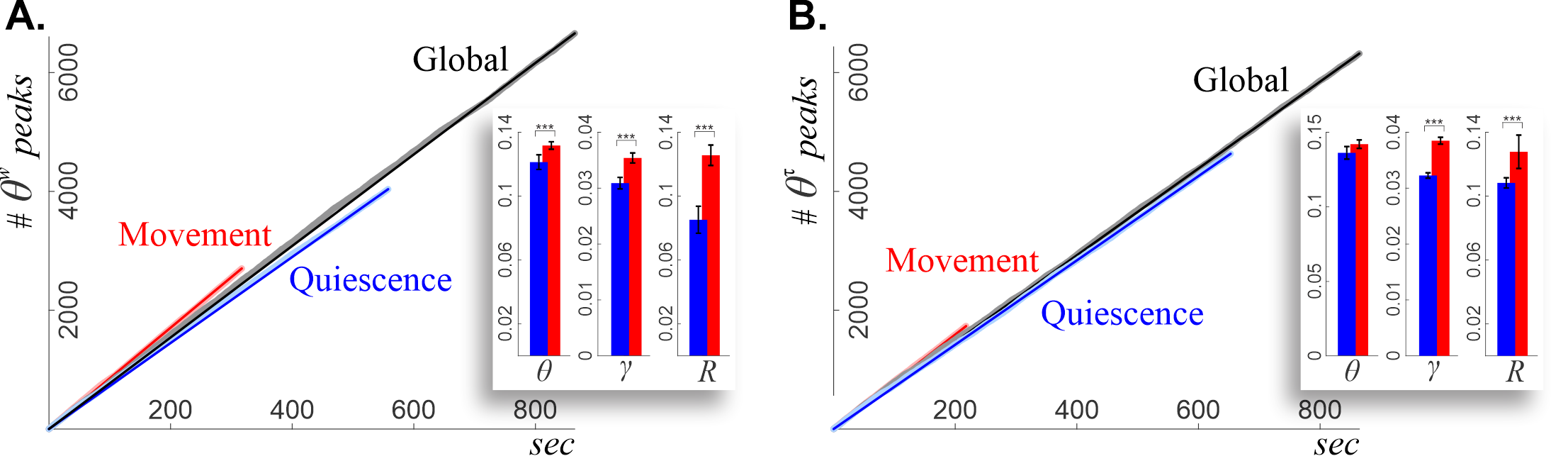}
	\caption{{\footnotesize
			\textbf{Brain wave trends}. Brain waves are oscillatory in nature, producing linear trends which
			can	be used as the basis for calculating $\lambda$. For example, the $\theta$ wave produces peaks
			that appear at frequency $\approx8$ Hz. However, the behavior of the animals may situationally 
			affect the expected patterning of brain rhythms. Correspondingly, computing $\lambda$ relative
			to specific means, i.e. for different physiological contexts, allows gaining a comprehensive 
			understanding of data haphazardness. 
			\textbf{A}. Globally averaged WT $\theta$-cadence amounts to approximately $8$ Hz. Yet, this value 
			changes in quiescence and in movement. The insert reveals significant differences in the $\theta$, 
			$\gamma$, and $R$ trends between movement and quiescence. 
			\textbf{B}. Brain wave frequencies of tau-mice also discriminate between activity and rest.
			Note that these differences are diminished in $R$ and are eliminated in $\theta$. 
	}}
	\label{fig:supp2}
	%\end{wrapfigure}
\end{figure}
%%%%%%%%%%%%%%%%%%%%%%%%%%%%%%%%%%

%%%%%%%%%%%%%%%%%%%%%%%%%%%%%%%%%%%%%%%
% SUPPLEMENTARY FIGURE 2
\renewcommand{\figurename}{Fig}
\begin{figure}[H]
	%\begin{wrapfigure}{c}{0.7\textwidth}
	\centering
	\includegraphics[scale=.9]{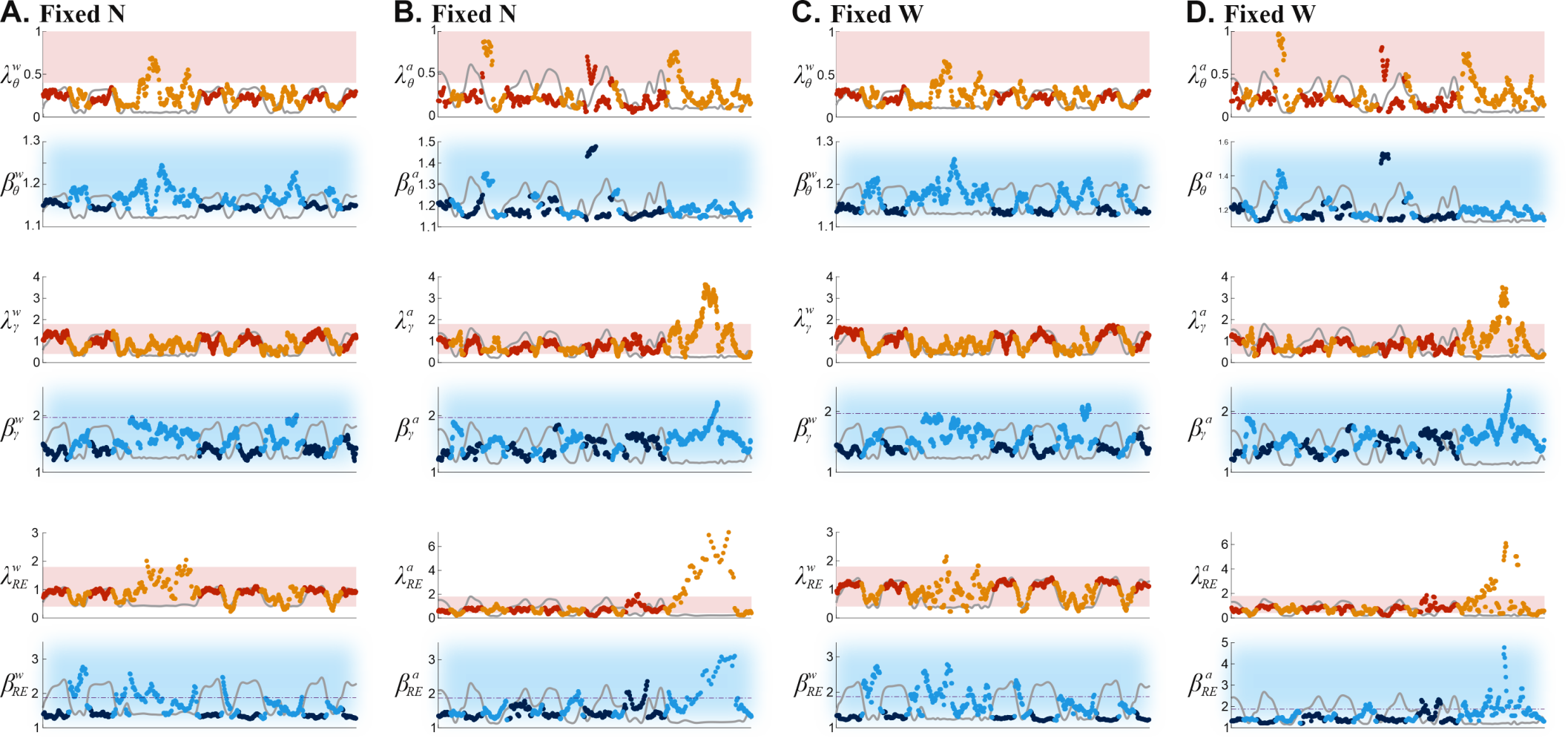}
	\caption{{\footnotesize
			\textbf{Fixed number of peaks per window versus fixed window size}. This figure illustrates the 
			difference in stochasticity dynamics evaluated using a fixed number of peaks per window as compared to 
			fixed time windows.
			\textbf{A}. Time window of fixed width, $n$, for WT mice. $n^w_{\theta}=16$, $n^w_{\gamma}=59$, and $n^w_{R}=15$ peaks. 
			\textbf{B}. Time window of fixed width, $n$, for tau-mice. $n^{\tau}_{\theta}=16$, $n^{\tau}_{\gamma}=57$, and $n^{\tau}_{R}=15$ peaks.
			\textbf{C}. Fixed time step, $L$, for WT mice. $L=2$ sec for all waves. 
			\textbf{D}. Fixed time step, $L$, for tau-mice. $L=2$ sec for all waves. 
	}}
	\label{fig:FixedN}
	%\end{wrapfigure}
\end{figure}
%%%%%%%%%%%%%%%%%%%%%%%%%%%%%%%%%%

%%%%%%%%%%%%%%%%%%%%%%%%%%%%%%%%%%%%%%%
%SUPPLEMENTAL FIGURE #3
\renewcommand{\figurename}{Fig}
\begin{figure}[H]
%	\begin{wrapfigure}{c}{0.7\textwidth}
	\centering
	\includegraphics[scale=1]{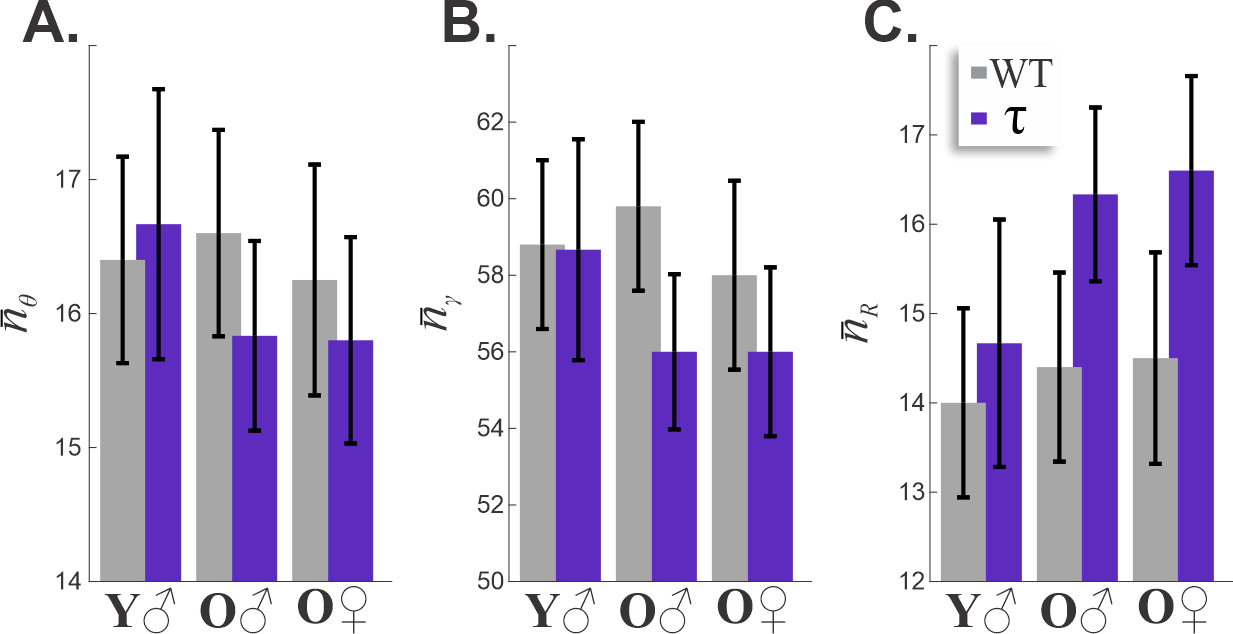}
	\caption{{\footnotesize
			\textbf{Average number of peaks in each wave, per two second window} is similar for all ages and phenotypes.  
			There were on average $n_{\theta}=16.2$ peaks of the $\theta$-wave (panel \textbf{A}), $n_{\gamma}=57.8$ peaks of 
			$\gamma$ (in panel \textbf{B}), and $n_{R}=15.2$ peaks (panel \textbf{C}) per window. 
	}}
	\label{fig:MeanNumPeaks}
%	\end{wrapfigure}
\end{figure}
%%%%%%%%%%%%%%%%%%%%%%%%%%%%%%%%%%

%%%%%%%%%%%%%%%%%%%%%%%%%%%%%%%%%%%%%%%
%SUPPLEMENTAL FIGURE 4
\renewcommand{\figurename}{Fig}
\begin{figure}[H]
	%\begin{wrapfigure}{c}{0.7\textwidth}
	\centering
	\includegraphics[scale=.94]{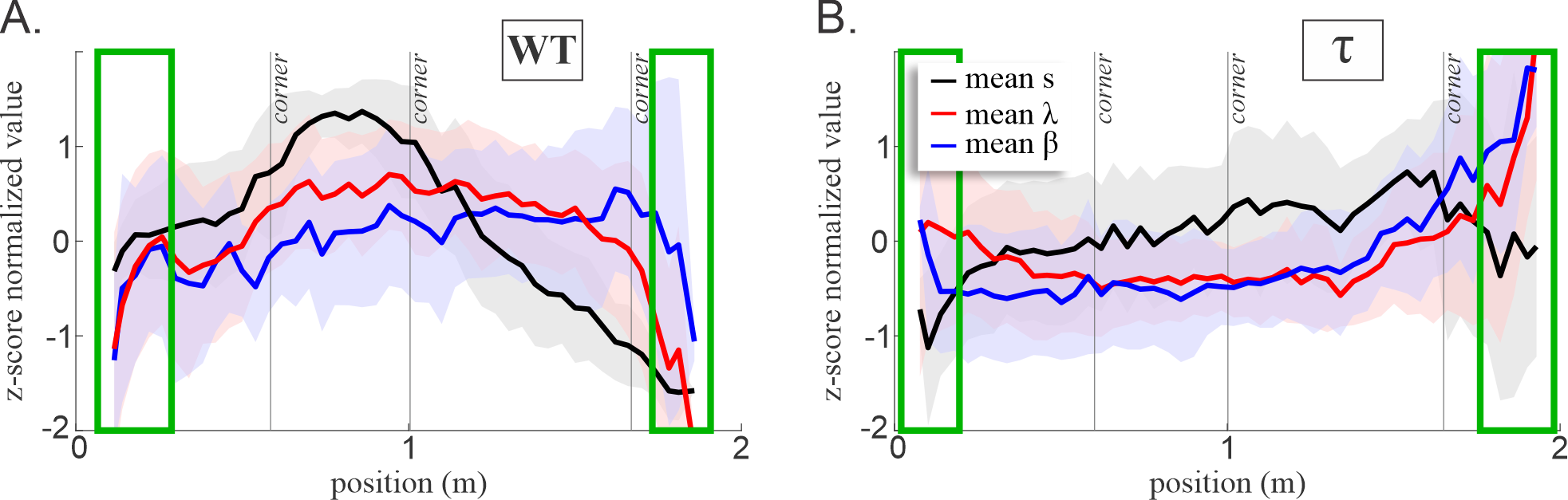}
	\caption{{\footnotesize
			\textbf{Mean stochasticity and speed along the mice's linearized trajectory}, with the corner
			locations demarcated by vertical lines. Standard deviations for all variables (color-coded) are
			represented by the shaded bands around the corresponding means. The speed and the $\lambda/
			\beta$-values for each lap were normalized by $z$-score and averaged.  
			\textbf{A}. WT mice produce LFP patterns that are not only consistent between laps, but 
			also appear to trend with the animals' speed. Note that $\lambda$, $\beta$, and speed correlate 
			at the ends of the trajectory (green boxes). 
			\textbf{B}. tau-mice scores, $\lambda^{\tau}$ and $\beta^{\tau}$ may oppose the speed, 
			increasing at food wells while speed decreases, and vice versa (green boxes). 
 	}}
	\label{fig:lamDescent}
	%\end{wrapfigure}
\end{figure}
%%%%%%%%%%%%%%%%%%%%%%%%%%%%%%%%%%

%%%%%%%%%%%%%%%%%%%%%%%%%%%%%%%%%%%%%%%
%SUPPLEMENTAL FIGURE 5
\renewcommand{\figurename}{Fig}
\begin{figure}[H]
	%\begin{wrapfigure}{c}{0.7\textwidth}
	\centering
	\includegraphics[scale=1.16]{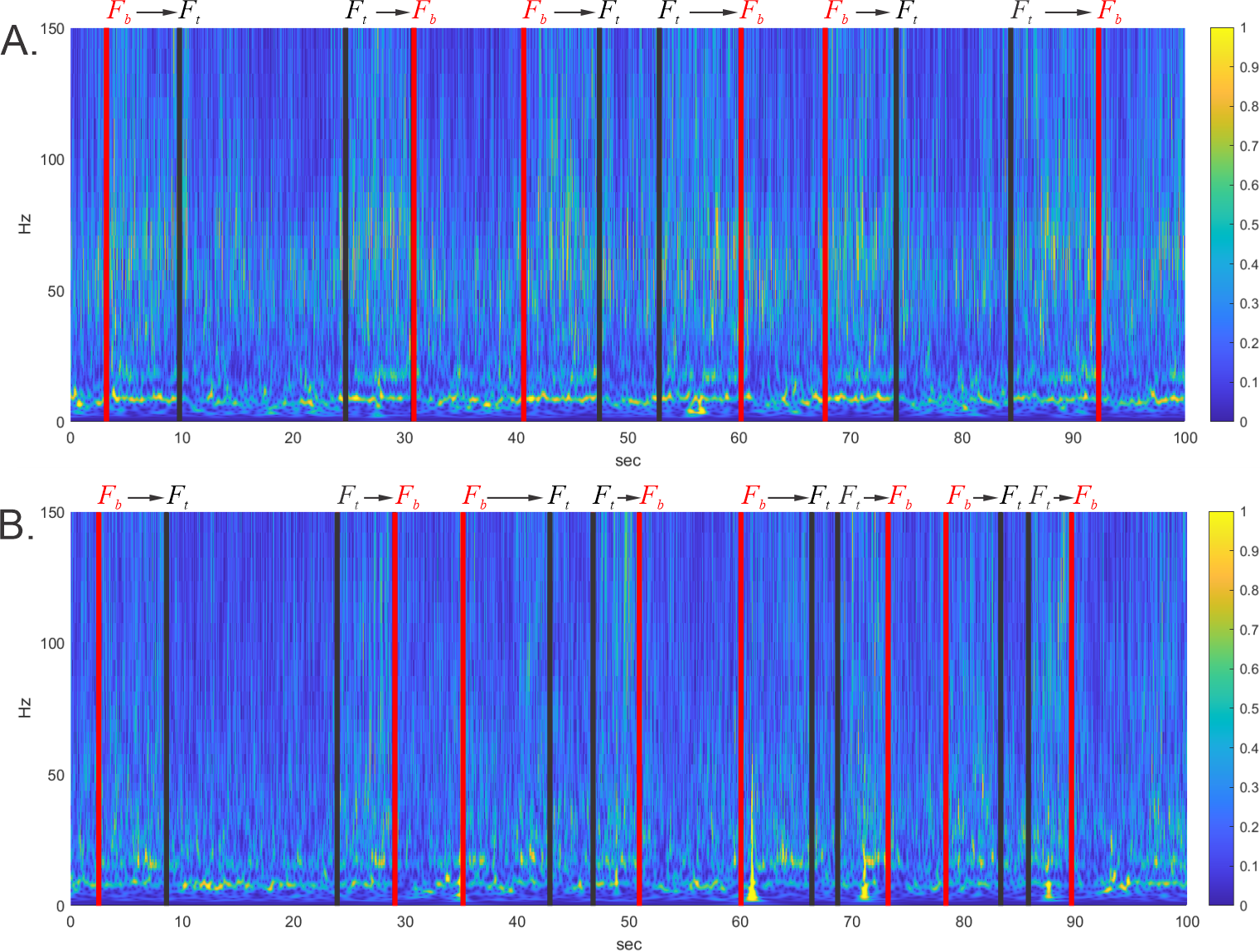}
	\caption{{\footnotesize
			\textbf{Continuous wavelet transform} of LFPs recorded from WT and tau-mice (panels \textbf{A}
			and \textbf{B}). Red lines marked by ``$F_b$" correspond to the bottom food well, and the black 
			lines marked by ``$F_t$" represent the top food well on Fig.~\ref{fig:Laps}. The spectrogram reveals
			minor changes in the power of the LFP frequencies as the mice change between active movement and
			quiescence, whereas pattern differences are dramatic (Figs.~\ref{fig:ThStoch}, \ref{fig:GmStoch},
			\ref{fig:RipStoch}).
	}}
	\label{fig:spec}
	%\end{wrapfigure}
\end{figure}

%%%%%%%%%%%%%%%%%%%%%%%%%%%%%%%%%%

%%%%%%%%%%%%%%%%%%%%%%%%%%%%%%%%%%%%%%%
%SUPPLEMENTAL FIGURE 6
\renewcommand{\figurename}{Fig}
\begin{figure}[H]
	%\begin{wrapfigure}{c}{0.45\textwidth}
	\centering
	\includegraphics[scale=.72]{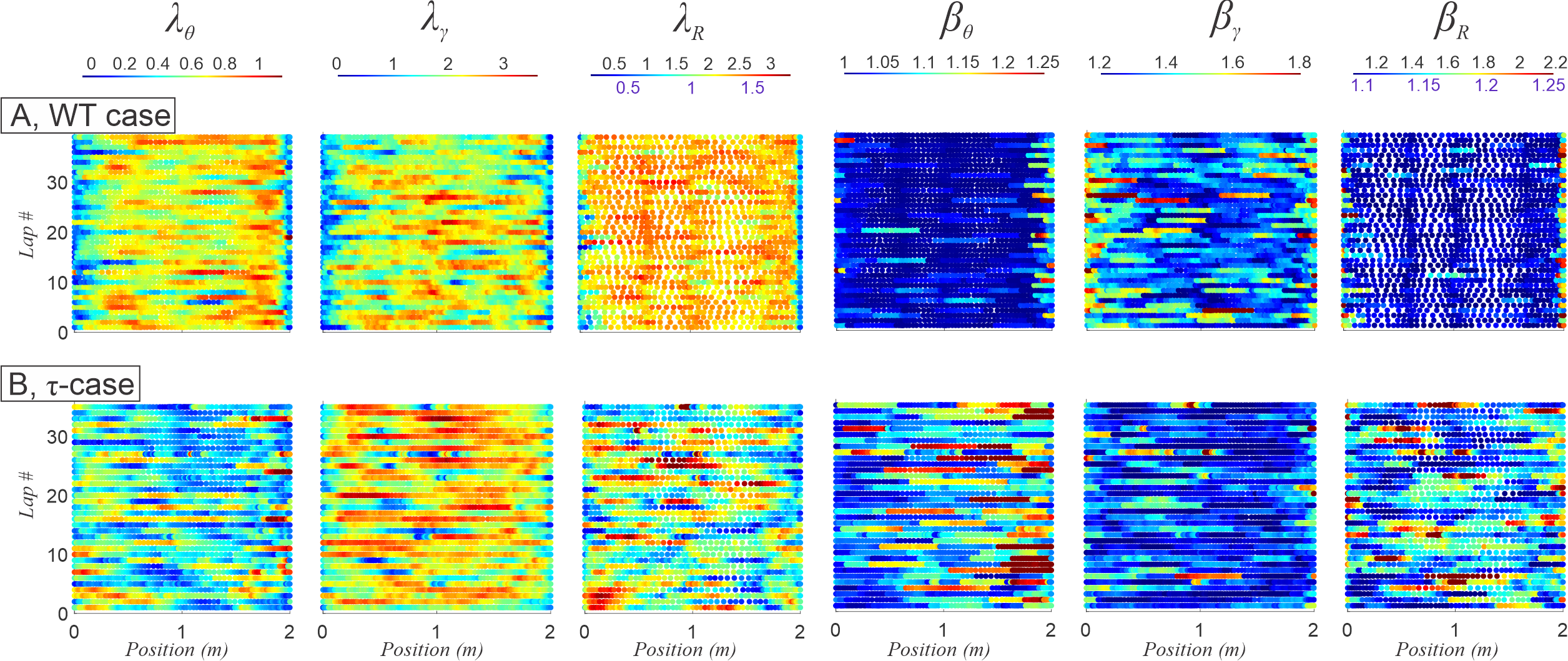}
	\caption{{\footnotesize
			\textbf{Linearized spatial stochasticity map of brainwaves by lap}. 
			\textbf{A}. The $\theta$, $\gamma$, and $R$ maps show consistent structures between laps
			in healthy mice.
			\textbf{B}. In tau-mice, maps are shuffled and spatial organization is lost. 
	}}
	\label{fig:Laps}
	%\end{wrapfigure}
\end{figure}
%%%%%%%%%%%%%%%%%%%%%%%%%%%%%%%%%%

%%%%%%%%%%%%%%%%%%%%%%%%%%%%%%%%%%%%%%%
%SUPPLEMENTAL FIGURE 7
\begin{figure}[H]
	\centering
	\includegraphics[scale=0.8]{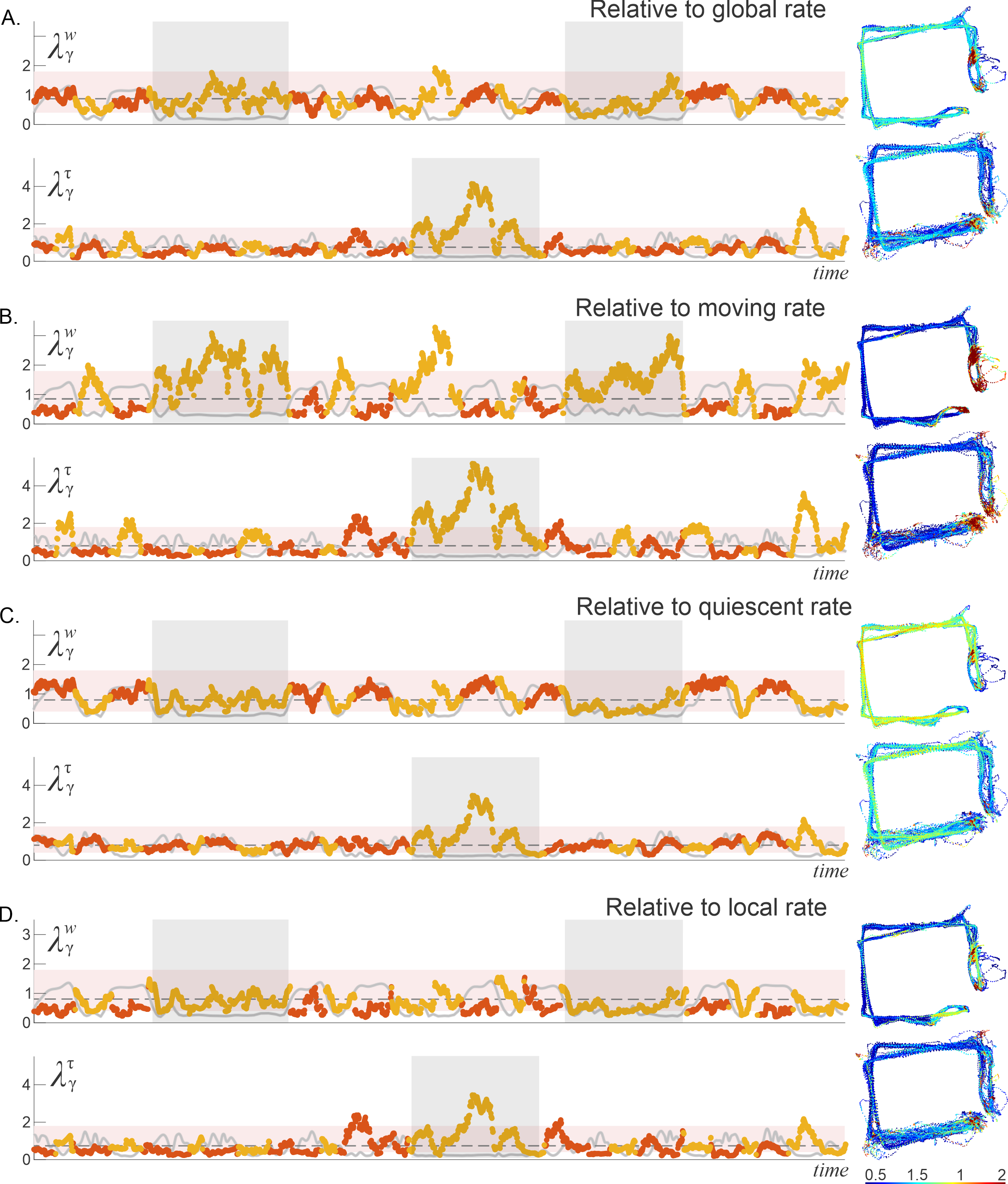}
	\caption{{\footnotesize
			\textbf{Haphazardness of $\gamma$-wave referenced to different physiological states in WT and 
				tau-mice}.
			\textbf{A}. Relative to the global average, $\gamma$-rhythms are more stochastic than
			$\theta$-patterns. $\lambda$-scores for WT (top) and tau (bottom) falls within the pink ``typicality"
			zone. Nevertheless, $\gamma$-patterns are sensitive to changes between movement and quiescence,
			rising during runs and falling during brief pit-stops at the food wells. $\lambda_\gamma$-maps on
			the right reveal similar spatial organization in both phenotypes. 
			\textbf{B}. Relative to the active track running, $\lambda_\gamma$ drops to atypical values during
			movements, indicating the designed (hence stochastically abnormal) adherence to the chosen trend,
			as well as rises to atypically high values during periods of prolonged quiescence. This is another
			``statistical perspective" on the qualitative differences between $\gamma$-cadence during quiescence
			and activity. Note the arrangement of the corresponding low and high $\lambda$-values on the spatial
			$\lambda$ maps (right). 
			\textbf{C}. Relative to quiescence, $\lambda_\gamma$ tends to follow the profile of speed. 
			\textbf{D}. Relative to the ongoing trend, $\lambda_\gamma$ remains low.  
			\textbf{E}. $\beta_\gamma$ for WT and tau-mice shown for reference. 
			Compare this to Fig.~\ref{fig:frame}.
	}}
	\label{fig:GmSlopes}
\end{figure}
%%%%%%%%%%%%%%%%%%%%%%%%%%%%%%%%%%

%%%%%%%%%%%%%%%%%%%%%%%%%%%%%%%%%%%%%%%
%SUPPLEMENTAL FIGURE 8
\begin{figure}[H]
	\centering
	\includegraphics[scale=1]{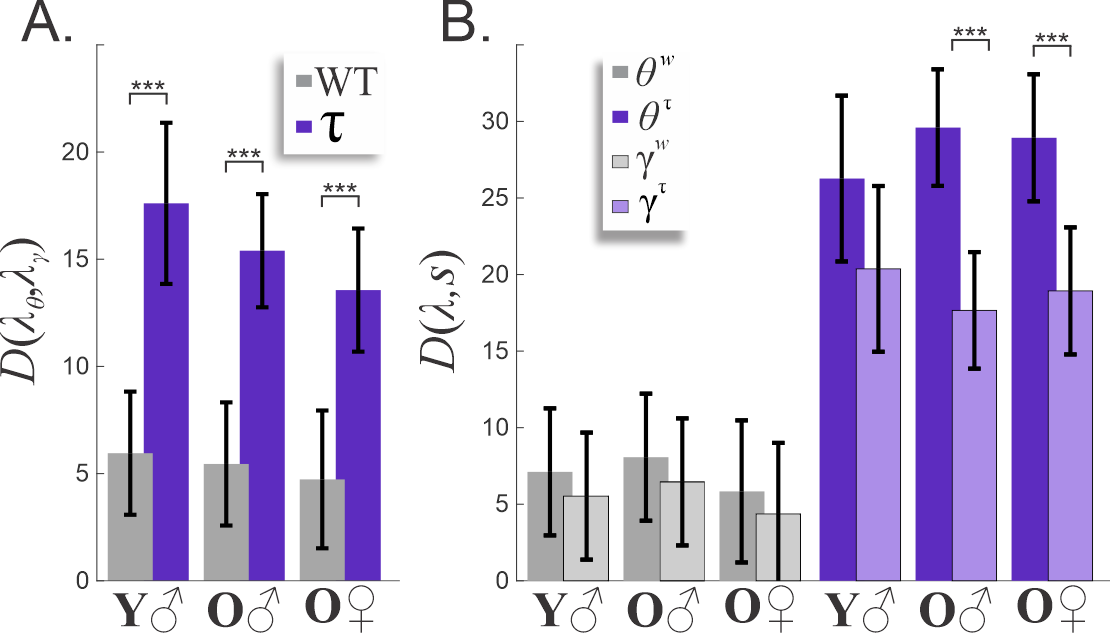}
	\caption{{\footnotesize
			\textbf{Comparing patterns of $\theta$ and $\gamma$}. 
			\textbf{A}. There is a significant increase ($\sim3\times$) in the LCSS distance between $\lambda_
			\theta$ and $\lambda_\gamma$ in tau-mice, regardless of the mice's age group or gender. 
			\textbf{B}. We compared LCSS distance between $\lambda_\theta$ or $\lambda_\gamma$ and the mouse's
			speed for each age group (old vs young) and gender (male vs female). The results reveal statistically
			significant distancing of speed from $\lambda_\theta$ and $\lambda_\gamma$ in old tau-mice---for them
			there exists a behavioral de-coupling from $\theta$- and $\gamma$-rhythmicity.
	}}
	\label{fig:ThVGm}
\end{figure}
%%%%%%%%%%%%%%%%%%%%%%%%%%%%%%%%%%

%%%%%%%%%%%%%%%%%%%%%%%%%%%%%%%%%%%%%%%
%SUPPLEMENTAL FIGURE 9
\renewcommand{\figurename}{Fig}
\begin{figure}[H]
	%\begin{wrapfigure}{c}{0.7\textwidth}
	\centering
	\includegraphics[scale=0.9]{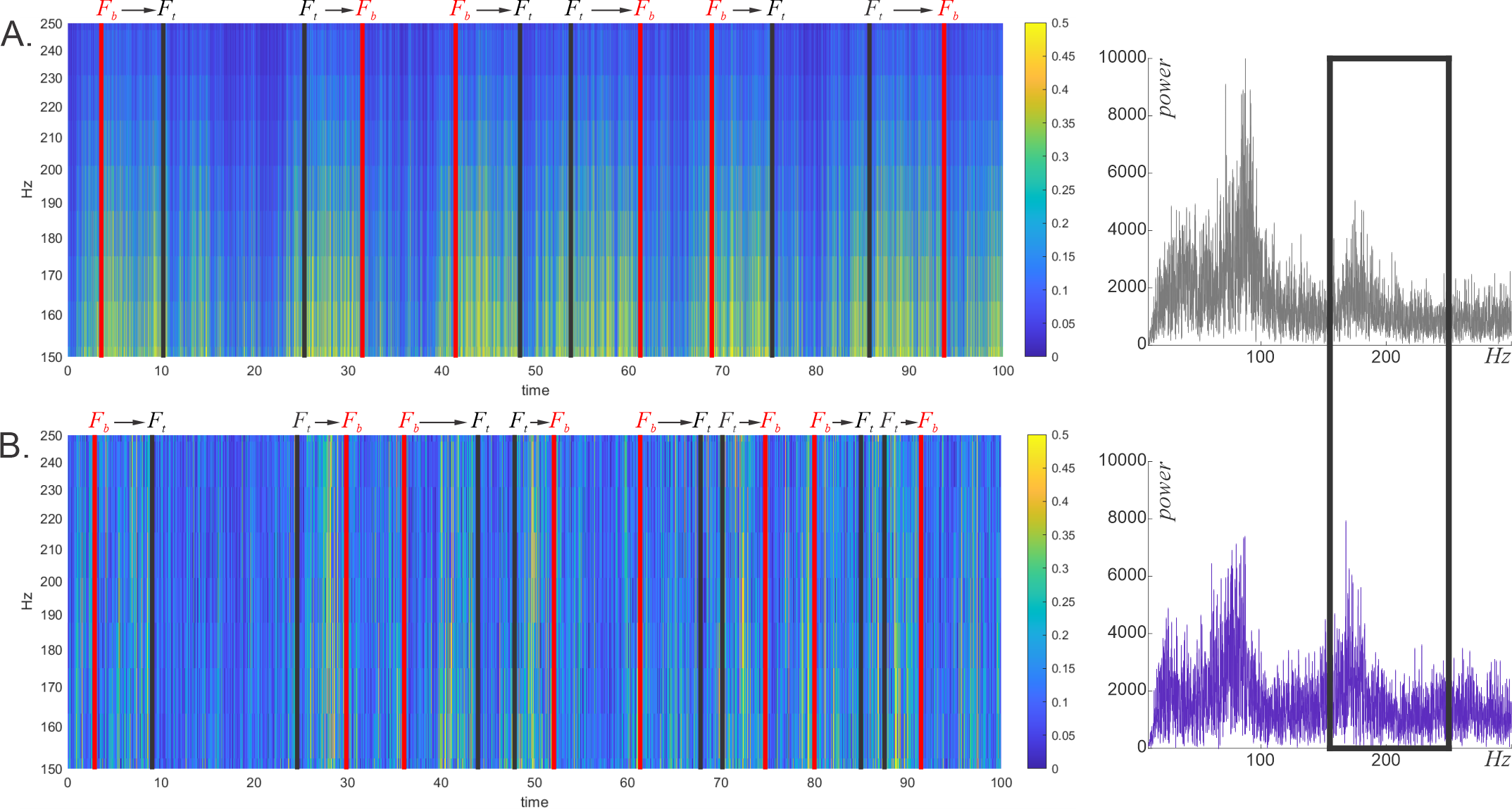}
	\caption{{\footnotesize
			\textbf{Comparative analysis for high frequency domains}.
			\textbf{A}. Continuous wavelet transform comparing high frequency ranges of LFPs recorded from WT
			(top) and tau (bottom) mice. Vertical red lines marked ``$F_b$" and black lines marked ``$F_t$" 
			correspond to the bottom and top food wells on Fig.~\ref{fig:Laps}. 
			\textbf{B}. The fast Fourier transforms for each phenotype and the boxed frequency range capture
			small differences between WT and tau-rippling, whereas pattern differences are much more explicit
			(Fig.~\ref{fig:RipStoch}).
	}}
	\label{fig:spechigh}
	%\end{wrapfigure}
\end{figure}
%%%%%%%%%%%%%%%%%%%%%%%%%%%%%%%%%%

%%%%%%%%%%%%%%%%%%%%%%%%%%%%%%%%%%%%%%%
%SUPPLEMENTAL FIGURE 10
\begin{figure}[H]
	\centering
	\includegraphics[scale=0.8]{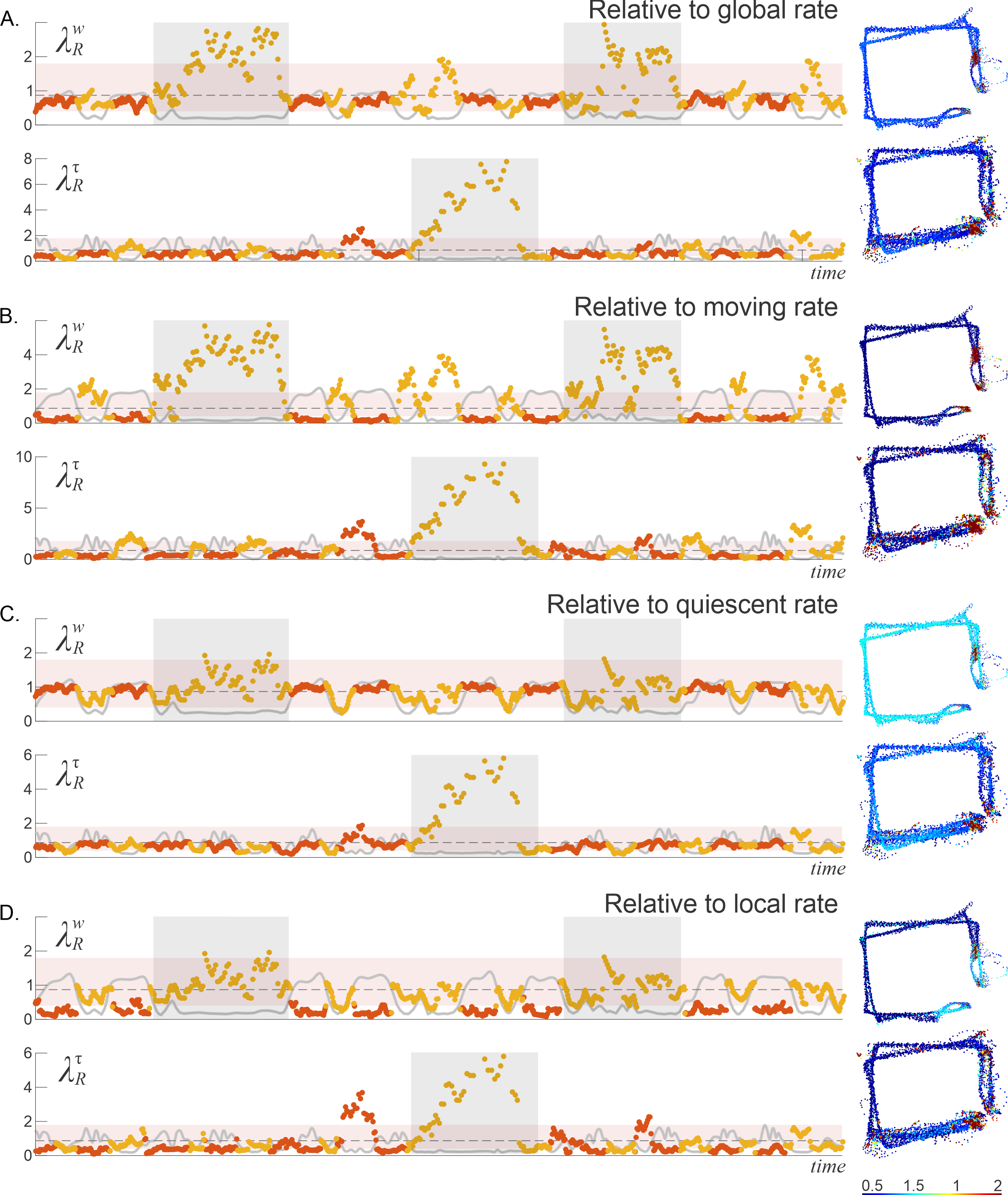}
	\caption{{\footnotesize
			\textbf{Haphazardness of ripple events referenced to different physiological states in WT and 
				tau-mice}.
			\textbf{A}. $R$-patterns calculated for WT (top) and tau (bottom) mice with reference to their
			respective global rates. WT ripple events are sensitive to changes in movement and quiescence,
			rising during runs and falling during rest. In contrast, $\lambda_{R}^{\tau}$ dissociates from
			the	speed. Spatial	maps of $\lambda_{R}^w$ show environmentally specific $R$-patterning and
			decreased environmental dependence in tau-mice. 
			\textbf{B}. Relative to the activity rate, WT rippling during movement is abnormally adherent
			to the expected trend, deviating from it during quiescence, while tau-ripples remain as 
			haphazard as they were relative to the global rate, as illustrated by the $\lambda_{R}^w$ and 
			$\lambda_{R}^{\tau}$ maps on the right.
			\textbf{C}. Relative to quiescence, $\lambda_{R}^w$ follows the profile of mouse's speed, whereas
			$\lambda_{R}^{\tau}$ remains unchanged. The spatial maps confirm the sensitivity of 
			$R^w$-stochasticity	to changes in the reference trend. 
			\textbf{D}. Haphazardness of $R$s referenced to the animal's ongoing behavior. 
	}}
	\label{fig:RpSlopes}
\end{figure}

%%%%%%%%%%%%%%%%%%%%%%%%%%%%%%%%%%

\newpage
\clearpage
\section{References}
\label{sec:ref}


\begin{thebibliography}{99}
	%1-5
	\bibitem{DeTure} DeTure, M., \& Dickson, D. The neuropathological diagnosis of Alzheimer’s disease. \textit{Mol Neurodegen}. \textbf{14}(1): 32 (2019).
	\bibitem{Perl} Perl, D. Neuropathology of Alzheimer's disease. \textit{Mt Sinai J Med}. \textbf(77)(1): 32-42 (2010).
	\bibitem{Wenk} Wenk, G. Neuropathologic changes in Alzheimer's disease. \textit{J Clin Psychiatry}. \textbf{64} Suppl 9: 7-10 (2003).
	\bibitem{Targa} Targa Dias Anastacio, H., Matosin, N., \& Ooi, L. Neuronal hyperexcitability in Alzheimer's disease: what are the drivers behind this aberrant phenotype? \textit{Translational psychiatry}, \textbf{12}(1): 257. (2022). %https://doi.org/10.1038/s41398-022-02024-7
	\bibitem{Bachmann} Bachmann, C., Tetzlaff, T., Duarte, R., \& Morrison, A. Firing rate homeostasis counteracts changes in stability of recurrent neural networks caused by synapse loss in Alzheimer's disease. \textit{PLoS Comput. Bio}., \textbf{16}(8): e1007790 (2020). %https://doi.org/10.1371/journal.pcbi.1007790
	
	%6-10
	\bibitem{Glenner} Glenner, G., Wong, C. Alzheimer's disease: initial report of the purification and characterization of a novel cerebrovascular amyloid protein. \textit{Biochem Biophys Res Commun}. \textbf{120}(3): 885-90 (1984).	
	\bibitem{Alzheimer} Alzheimer, A., Stelzmann, R., Schnitzlein, H., \& Murtagh, F. An English translation of Alzheimer's 1907 paper, \textit{Uber eine eigenartige Erkankung der Hirnrinde}. \textit{Clin Anat}. \textbf{8}(6): 429-31 (1995).
	\bibitem{Oddo} Oddo, S., Caccamo, A., Kitazawa, M., Tseng, B., LaFerla, F. Amyloid deposition precedes tangle formation in a triple transgenic model of Alzheimer's disease. \textit{Neurobiol Aging}. \textbf{24}(8): 1063-70 (2003).
	\bibitem{Selkoe2} Selkoe, D. Alzheimer's disease: genes, proteins, and therapy. \textit{Physiol Rev}. \textbf{81}(2): 741-66 (2001).
	\bibitem{Martinsson} Martinsson, I., Quintino, L., Garcia, M., Konings, S., Torres-Garcia, L., Svanbergsson, A., Stange, O., England, R., Deierborg, T., Li, J., Lundberg, C., Gouras, G. A$\beta$/Amyloid Precursor Protein-Induced Hyperexcitability and Dysregulation of Homeostatic Synaptic Plasticity in Neuron Models of Alzheimer’s Disease. \textit{Front Aging Neurosci}. \textbf{14}: 946297 (2022).
	
	%11-15
	\bibitem{Frere} Frere, S., Slutsky, I. Alzheimer's Disease: From Firing Instability to Homeostasis Network Collapse. \textit{Neuron}. \textbf{97}(1): 32-58 (2018). 
	\bibitem{Rae-Grant} Rae-Grant, A., Blume, W., Lau, C., Hachinski, V., Fisman, M., \& Merskey, H. The Electroencephalogram in Alzheimer-Type Dementia: A Sequential Study Correlating the Electroencephalogram With Psychometric and Quantitative Pathologic Data. \textit{Archives of Neurology}. \textbf{44}(1): 50-4 (1987).
	\bibitem{Liddell} Liddell, D. Investigations of E.E.G. findings in presenile dementia. \textit{J Neurol Neurosurg Psychiatry}. \textbf{21}(3): 173-6 (1958).
	\bibitem{Jelles} Jelles, B., van Birgelen, J., Slaets, J., Hekster, R., Jonkman, E., \& Stam, C. Decrease of non-linear structure in the EEG of Alzheimer patients compared to healthy controls. \textit{Clin Neurophys}. \textbf{110}(7): 1159-67 (1999).
	\bibitem{Palop} Palop, J., Chin, J., Roberson, E., Wang, J., Thwin, M., Bien-Ly, N., Yoo, J., Ho, K., Yu, G., Kreitzer, A., Finkbeiner, S., Noebels, J., Mucke, L. Aberrant excitatory neuronal activity and compensatory remodeling of inhibitory hippocampal circuits in mouse models of Alzheimer's disease. \textit{Neuron}. \textbf{55}(5): 697-711 (2007).
	
	%16-20
	\bibitem{Born} Born, H., Kim, J., Savjani, R., Das, P., Dabaghian, Y., Guo, Q., Yoo, J., Schuler, D., Cirrito, J., Zheng, H., Golde, T., Noebels, J., Jankowsky, J. Genetic suppression of transgenic APP rescues Hypersynchronous network activity in a mouse model of Alzeimer's disease. \textit{J Neurosci}. \textbf{34}(11): 3826-40 (2014).
	\bibitem{Scarmeas} Scarmeas, N., Honig, L., Choi, H., Cantero, J., Brandt, J., Blacker, D., Albert, M., Amatniek, J., Marder, K., Bell, K., Hauser, W., Stern, Y. Seizures in Alzheimer disease: who, when, and how common? \textit{Arch Neurol}. \textbf{66}(8): 992-7 (2009). 
	\bibitem{Palop2} Palop, J., Mucke, L. Network abnormalities and interneuron dysfunction in Alzheimer disease. \textit{Nat Rev Neurosci}. \textbf{17}(12): 777-792 (2016).
	\bibitem{Pandis} Pandis, D., Scarmeas, N. Seizures in Alzheimer disease: clinical and epidemiological data. \textit{Epilepsy Curr}. \textbf{12}(5): 184-7 (2012). 
	\bibitem{Vossel} Vossel, K., Ranasinghe, K., Beagle, A., Mizuiri, D., Honma, S., Dowling, A., Darwish, S., Van Berlo, V., Barnes, D., Mantle, M., Karydas, A., Coppola, G., Roberson, E., Miller, B., Garcia, P., Kirsch, H., Mucke, L., Nagarajan, S. Incidence and impact of subclinical epileptiform activity in Alzheimer's disease. \textit{Ann Neurol}. \textbf{80}(6): 858-870 (2016).
	
	%21-25
	\bibitem{Hoffman} Hoffman, C., Cheng, J., Ji, D., Dabaghian, Y. Pattern dynamics and stochasticity of the brain rhythms. \textit{Proc. Natl. Acad. Sci}. \textbf{120}(14): e2218245120 (2023). 
	
	\bibitem{Ramsden} Ramsden, M., Kotilinek, L., Forster, C., Paulson, J., McGowan, E., SantaCruz, K., Guimaraes, A., Yue, M., Lewis, J., Carlson, G., Hutton, M., \& Ashe, K. Age-dependent neurofibrillary tangle formation, neuron loss, and memory impairment in a mouse model of human tauopathy (P301L). \textit{J Neurosci}. \textbf{25}(46): 10637-47 (2005).
	\bibitem{SantaCruz} SantaCruz, K., Lewis, J., Spires, T., Paulson, J., Kotilinek, L., Ingelsson, M., Guimaraes, A., DeTure, M., Ramsden, M., McGowan, E., Forster, C., Yue, M., Orne, J., Janus, C., Mariash, A., Kuskowski, M., Hyman, B., Hutton, M., \& Ashe, K. Tau Suppression in a Neurodegenerative Mouse Model Improves Memory Function. \textit{Science}. \textbf{309}(5733): 476-81 (2005).
	
	\bibitem{ChengJi} Cheng, J. \& Ji, D. Rigid firing sequences undermine spatial memory codes in a neurodegenerative mouse model. \textit{eLife}, \textbf{2}: e00647 (2013). %https://doi.org/10.7554/eLife.00647
	\bibitem{Ciupek} Ciupek, S., Cheng, J., Ali, Y., Lu, H., \& Ji, D. Progressive functional impairments of hippocampal neurons in a tauopathy mouse model. \textit{J. Neurosci}. \textbf{35}(21): 8118–8131 (2015). 
	
	\bibitem{Kolm} Kolmogorov, A. Sulla determinazione empirica di una legge di distribuzione. \textit{Giornale dell'Istituto Italiano degli Attuari,} \textbf{4}(1): 83-91 (1933).
	\bibitem{Arn1} Arnold, V. Orbits' statistics in chaotic dynamical systems. \textit{Nonlinearity} \textbf{21}: T109 (2008).
	\bibitem{Arn2} Arnold, V. Empirical study of stochasticity for deterministic chaotic dynamics of geometric progressions of residues. \textit{Funct. Anal. Other Math}. \textbf{2}: 139-149 (2009).
	\bibitem{Arn3} Arnold, V. To what extent are arithmetic progressions of fractional parts stochastic? \textit{Russian Mathematical Surveys} \textbf{63}: 205 (2008).
	
	%26-30
	\bibitem{Rickert} Rickert, J., Oliveira, S., Vaadia, E., Aertsen, A., Rotter, S., Mehring, C. Encoding of movement direction in different frequency ranges of motor cortical local field potentials. \textit{J Neurosci}. \textbf{25}(39): 8815-24 (2005). 
	\bibitem{Baker} Baker, S., Kilner, J., Pinches, E., Lemon, R. The role of synchrony and oscillations in the motor output. \textit{Exp Brain Res}. \textbf{128}(1-2): 109-17 (1999).
	\bibitem{Murthy} Murthy, V., Fetz, E. Oscillatory activity in sensorimotor cortex of awake monkeys: synchronization of local field potentials and relation to behavior. \textit{J Neurophysiol}. \textbf{76}(6): 3949-67 (1996).
	
	%31-35
	\bibitem{BuzTheta1} Buzs\'aki, G. Theta oscillations in the hippocampus. \textit{Neuron} \textbf{33}: 325-340 (2002).
	\bibitem{Burgess} Burgess, N. \& O'Keefe, J. The theta rhythm. \textit{ Hippocampus} \textbf{15}(7): 825-826 (2005).
	\bibitem{BuzTheta2}	Buzs\'aki, G. Theta rhythm of navigation: link between path integration and landmark navigation, episodic and semantic memory. \textit{Hippocampus}. \textbf{15}(7): 827-40 (2005). 
	\bibitem{BuzTheta3}	Buzs\'aki, G., Moser, E. Memory, navigation and theta rhythm in the hippocampal-entorhinal system. \textit{Nat. Neurosci}. \textbf{16}(2): 130-8 (2013).
	\bibitem{Cacucci} Cacucci, F., Lever, C., Wills, T., Burgess, N., O'Keefe, J. Theta-modulated place-by-direction cells in the hippocampal formation in the rat. \textit{J. Neurosci}., \textit{24}(38): 8265-77 (2004).
	
	%36-40
	\bibitem{Itskov} Itskov, V., Pastalkova, E., Mizuseki, K., Buzs\'aki, G., Harris, K. Theta-mediated dynamics of spatial information in hippocampus. \textit{J. Neurosci}., \textbf{28}(23): 5959-64 (2008).
	\bibitem{Osbert} Osbert, C., Berj, L. Theta phase precession and phase selectivity: a cognitive device description of neural coding. \textit{J Neural Eng}. \textbf{6}(3): 036002 (2009).
	\bibitem{Jelic} Jelic, V., Johansson, S., Almkvist, O., Shigeta, M., Julin, P., Nordberg, A., Winblad, B., Wahlund, L. Quantitative electroencephalography in mild cognitive impairment: longitudinal changes and possible prediction of Alzheimer’s disease. \textit{Neurobiology of Aging}. \textbf{21}(4): 533-40 (2000). 
	\bibitem{Bennys} Bennys, K., Rondouin, G., Vergnes, C., Touchon, J. Diagnostic value of quantitative EEG in Alzheimer's disease. \textit{Neurophysiol Clin}. \textbf{31}(3): 153-60 (2001).
	\bibitem{ColginTh} Colgin, L. Mechanisms and functions of theta rhythms. \textit{Ann. Rev. Neurosci}., \textbf{36}:295–312 (2013).
	
	%41-45
	\bibitem{Joshi} Joshi, A., Denovellis, E., Mankili, A., Meneksedag, Y., Davidson, T., Gillespie, A., Guidera, J., Roumis, D., Frank, L. Dynamic synchronization between hippocampal representations and stepping. \textit{Nature}. \textbf{617}(7959): 125-131 (2023).
	\bibitem{Bender} Bender, F., Gorbati, M., Cadavieco, M., Denisova, N., Gao, X., Holman, C., Korotkova, T., Ponomarenko, A. Theta oscillations regulate the speed of locomotion via a hippocampus to lateral septum pathway. \textit{Nat Commun}. \textbf{6}: 8521 (2015). 
	\bibitem{Fuhrmann} Fuhrmann, F., Justus, D., Sosulina, L., Kaneko, H., Beutel, T., Friedrichs, D., Schoch, S., Schwarz, M., Fuhrmann, M., Remy, S. Locomotion, Theta Oscillations, and the Speed-Correlated Firing of Hippocampal Neurons Are Controlled by a Medial Septal Glutamatergic Circuit. \textit{Neuron}. \textbf{86}(5): 1253-64 (2015). 
	\bibitem{Vlachos} Vlachos, M., Kollios, G., \& Gunopulos, D. Discovering similar multidimensional trajectories. \textit{Proceedings 18th International Conference on Data Engineering}. 673-684 (2002).
	\bibitem{Khan} Khan, R., Ahmad, M., \& Zakarya, M. Longest Common Subsequence Based Algorithm for Measuring Similarity Between Time Series: A New Approach. \textit{World Applied Sciences Journal}. \textbf{24} (2013).
		
	%46-50
	\bibitem{Morse} Morse, M., Patel, J. An efficient and accurate method for evaluating time series similarity. \textit{ACM SIGMOD Conference}. (2007).
	\bibitem{OKeefe} O'Keefe, J., Dostrovsky, J. The hippocampus as a spatial map. Preliminary evidence from unit activity in the freely-moving rat. \textit{Brain Research}. \textbf{34}(1): 171–75 (1971).
	\bibitem{MosRev} Moser, E. Kropff, E. \& Moser, M.-B. Place Cells, Grid Cells, and the Brain's Spatial Representation System. \textit{Annu Rev Neurosci}. \textbf{31}(1): 69-89 (2008).
	\bibitem{Fenton} Fenton, A., Kao, H., Neymotin, S., Olypher, A., Vayntrub, Y., Lytton, W., Ludvig, N. Unmasking the CA1 ensemble place code by exposures to small and large environments: more place cells and multiple, irregularly arranged, and expanded place fields in the larger space. \textit{J Neurosci}. \textbf{28}(44): 11250-62 (2008). 
	\bibitem{Redish} Redish, A., Battaglia, F., Chawla, M., Ekstrom, A., Gerrard, J., Lipa, P., Rosenzweig, E., Worley, P., Guzowski, J., McNaughton, B., Barnes, C. Independence of firing correlates of anatomically proximate hippocampal pyramidal cells. \textit{J Neurosci}. \textbf{21}(5): RC134 (2001). 
		
	%51-55
	\bibitem{Amaral} Amaral, D. Emerging principles of intrinsic hippocampal organization. \textit{Curr Opin Neurobiol}. \textbf{3}(2): 225-9 (1993). 
	\bibitem{McNaughton} McNaughton, B., Barnes, C., Meltzer, J., Sutherland, R. Hippocampal granule cells are necessary for normal spatial learning but not for spatially-selective pyramidal cell discharge. \textit{Exp Brain Res}. \textbf{76}(3): 485-96 (1989). 
	\bibitem{Geisler} Geisler, C., Diba, K., Pastalkova, E., Mizuseki, K., Royer, S., Buzs\'aki, G. Temporal delays among place cells determine the frequency of population theta oscillations in the hippocampus. \textit{Proc. Natl. Acad. Sci}. \textbf{107}(17): 7957-62 (2010). 
	\bibitem{Dragoi} Dragoi, G., Buzs\'aki, G. Temporal encoding of place sequences by hippocampal cell assemblies. \textit{Neuron}. \textbf{50}(1): 145-57 (2006). 
	\bibitem{Jensen} Jensen, O., Lisman, J. Position reconstruction from an ensemble of hippocampal place cells: contribution of theta phase coding. \textit{J Neurophysiol}. \textbf{83}(5): 2602-9 (2000).
	
	%56-60
	\bibitem{Brown} Brown, E., Frank, L., Tang, D., Quirk, M., Wilson, M. A statistical paradigm for neural spike train decoding applied to position prediction from ensemble firing patterns of rat hippocampal place cells. \textit{J Neurosci}. \textbf{18}(18): 7411-25 (1998). 
	\bibitem{Foster} Foster, D., \& Wilson, M. Reverse replay of behavioural sequences in hippocampal place cells during the awake state
	\textit{Nature} \textbf{440}: 680–683 (2006).
	\bibitem{Karlsson} Karlsson, M. \& Frank L. Awake replay of remote experiences in the hippocampus. \textit{Nat. Neurosci}. \textbf{12}: 913-918 (2009).
	\bibitem{Carr} Carr, M., Jadhav, S. \& Frank, L. Hippocampal replay in the awake state: a potential substrate for memory consolidation and retrieval, \textit{Nat. Neurosci}., \textbf{14}: 147-153 (2011).
	\bibitem{Singer} Singer, A., Carr, M. Karlsson, M. \& Frank, L. Hippocampal SWR Activity Predicts Correct Decisions during the Initial Learning of an Alternation Task. \textit{Neuron} \textbf{77}: 1163-1173 (2013).
	
	%61-65
	\bibitem{Denovellis} Denovellis, E., Gillespie, A., Coulter, M., Sosa, M., Chung, J., Eden, U. \& Frank, L. Hippocampal replay of experience at real-world speeds. \textit{Elife} \textbf{10}:e64505 (2021).
	\bibitem{Nadasdy} N\'adasdy, Z., Hirase, H., Czurk\'o, A., Csicsvari, J., Buzs\'aki, G. Replay and time compression of recurring spike sequences in the hippocampus. \textit{J Neurosci}. \textbf{19}(21): 9497-507 (1999).
	\bibitem{Lee} Lee, A., Wilson, M. Memory of sequential experience in the hippocampus during slow wave sleep. \textit{Neuron}. \textbf{36}(6): 1183-94 (2002). 
	\bibitem{Ji} Ji, D. \& Wilson, M. Coordinated memory replay in the visual cortex and hippocampus during sleep, \textit{Nat. Neurosci}, \textbf{10}(1): 100–107 (2007).
	\bibitem{Diba} Diba, K., Buzs\'aki, G. Forward and reverse hippocampal place-cell sequences during ripples. \textit{Nat Neurosci}. \textbf{10}(10): 1241-2 (2007).
	\bibitem{Davidson} Davidson, T., Kloosterman, F., Wilson, M. Hippocampal replay of extended experience. \textit{Neuron}. \textbf{63}(4): 497-507 (2009).
	
	
	
	\bibitem{Gupta} Gupta, A., van der Meer, M., Touretzky, D. \& Redish, A. Hippocampal replay is not a simple function of experience. \textit{Neuron}, \textbf{65}(5): 695–705 (2010).
	\bibitem{Buhry} Buhry, L., Azizi, A., \& Cheng, S. Reactivation, replay, and preplay: how it might all fit together. \textit{Neural plasticity}, 203462 (2011). 
	\bibitem{Stella} Stella, F., Baracskay, P., O'Neill, J. \& Csicsvari, J. Hippocampal Reactivation of Random Trajectories Resembling Brownian Diffusion. \textit{Neuron}, \textbf{102}(2): 450–461.e7 (2019). 
	\bibitem{Kudrimoti} Kudrimoti, H., Barnes, C., \& McNaughton B. Reactivation of hippocampal cell assemblies: effects of behavioral state, experience, and EEG dynamics. \textit{J Neurosci}. \textbf{19}: 4090–101 (1999).
	
	%66-70
	\bibitem{ONeil} O'Neil, J., Senior, A., Allen, K., Huxter, J. \& Csicsvari, J. Reactivation of experience-dependent
	cell assembly patterns in the hippocampus. \textit{Nat. Neurosci}. \textbf{11}: 209–216 (2008).
	\bibitem{Olafsdottir} \'{O}lafsd\'{o}ttir, H., Bush, D. \& Barry, C. The Role of Hippocampal Replay in Memory and Planning. \textit{Current Biology}, \textbf{28}(1): R37-R50 (2018).
	\bibitem{Pastalkova} E. Pastalkova, V. Itskov, A. Amarasingham and G. Buzsaki. Internally Generated Cell Assembly Sequences in the Rat Hippocampus. \textit{Science} \textbf{321}: 1322 (2008).
	\bibitem{Taxidis} Taxidis, J., Anastassiou Costas, A., Diba, K. \& Koch, C. Local Field Potentials Encode Place Cell Ensemble Activation during Hippocampal Sharp Wave Ripples. \textit{Neuron}, \textbf{87}(3): 590-604 (2015).
	\bibitem{Ylinen} Ylinen, A., Bragin, A., N\'adasdy, Z., Jand\'o, G., Szab\'o, I., Sik, A., Buzs\'aki, G. Sharp wave-associated high-frequency oscillation (200 Hz) in the intact hippocampus: network and intracellular mechanisms. \textit{J Neurosci}. \textbf{15}(1 Pt 1): 30-46 (1995).
	
	%71-75
	\bibitem{Csi2} Csicsvari, J., Hirase, H., Czurk\'o, A., Mamiya, A., Buzs\'aki, G. Oscillatory coupling of hippocampal pyramidal cells and interneurons in the behaving Rat. \textit{J Neurosci}. \textbf{19}(1): 274-87 (1999). 
	\bibitem{Klaus1} Klausberger, T., Somogyi, P. Neuronal diversity and temporal dynamics: the unity of hippocampal circuit operations. \textit{Science}. \textbf{321}(5885): 53-7 (2008).
	\bibitem{Squire} Squire, L. Memory and the hippocampus: a synthesis from findings with rats, monkeys, and humans. \textit{Psychol Rev}. \textbf{99}(2): 195-231 (1992).
	\bibitem{Remondes} Remondes, M., Schuman, E. Role for a cortical input to hippocampal area CA1 in the consolidation of a long-term memory. \textit{Nature}. \textbf{431}(7009): 699-703 (2004).
	\bibitem{Manns} Manns, J., Zilli, E., Ong, K., Hasselmo, M, Eichenbaum, H. Hippocampal CA1 spiking during encoding and retrieval: relation to theta phase. \textit{Neurobiol Learn Mem}. \textbf{87}(1): 9-20 (2007).
	
	%76-80
	\bibitem{Moser} Moser, E., Kropff, E., Moser, M. Place cells, grid cells, and the brain's spatial representation system. \textit{Ann. Rev. Neurosci}. \textbf{31}: 69-89 (2008). 
	\bibitem{Klaus2} Klausberger, T., Magill, P., M\'arton, L., Roberts, J., Cobden, P., Buzs\'aki, G., Somogyi, P. Brain-state- and cell-type-specific firing of hippocampal interneurons in vivo. \textit{Nature}. \textbf{421}(6925): 844-8 (2003). 
	\bibitem{Jinno} Jinno, S., Klausberger, T., Marton, L., Dalezios, Y., Roberts, J., Fuentealba, P., Bushong, E., Henze, D., Buzs\'aki, G., Somogyi, P. Neuronal diversity in GABAergic long-range projections from the hippocampus \textit{J Neurosci}. \textbf{27}: 8790–8804 (2007).
	\bibitem{Varga} Varga, C., Golshani, P., Soltesz, I. Frequency-invariant temporal ordering of interneuronal discharges during hippocampal oscillations in awake mice. \textit{Proc. Natl. Acad. Sci}. \textbf{109}(40): E2726-34 (2012). 
	\bibitem{ClgMsr} Colgin, L., Denninger, T., Fyhn, M., Hafting, T., Bonnevie, T., Jensen, O., Moser, M-B. \& Moser, E. Frequency of gamma oscillations routes flow of information in the hippocampus. \textit{Nature} \textbf{462}: 353-357 (2009).
		
	%81-85
	\bibitem{Jia} Jia, X., Kohn, A. Gamma rhythms in the brain. \textit{PLoS Biol}. \textbf{9}(4): e1001045 (2011).
	\bibitem{Nikoli} Nikoli, D., Fries, P. \& Singer, W. Gamma oscillations: precise temporal coordination without a metronome. \textit{Trends Cogn Sci}. \textbf{17}: 54-55 (2013).
	\bibitem{Bieri} Bieri, K., Bobbitt, K., Colgin, L. Slow and Fast Gamma Rhythms Coordinate Different Spatial Coding Modes in Hippocampal Place Cells. \textit{Neuron}. \textbf{82}(3): 670-81 (2014).
	\bibitem{ColginGm} Colgin, L. \& Moser E. Gamma oscillations in the hippocampus. \textit{Physiology} \textbf{25}: 319-329 (2010).
	\bibitem{Goutagny} Goutagny, R., Gu, N., Cavanagh, C., Jackson, J., Chabot, J., Quirion, R., Krantic, S., Williams, S. Alterations in hippocampal network oscillations and theta-gamma coupling arise before A$\beta$ overproduction in a mouse model of Alzheimer's disease. \textit{European J Neurosci}. \textbf{37}(12): 1896-902 (2013).
	
	%86-90
	\bibitem{Iaccarino} Iaccarino, H., Singer, A., Martorell, A., Rudenko, A., Gao, F., Gillingham, T., Mathys, H., Seo, J., Kritskiy, O., Abdurrob, F., Adaikkan, C., Canter, R., Rueda, R., Brown, E., Boyden, E., Tsai, L. Gamma frequency entrainment attenuates amyloid load and modifies microglia. \textit{Nature}. \textbf{540}(7632): 230-235 (2016).  
	\bibitem{Mably} Mably, A., Gereke, B., Jones, D., Colgin, L. Impairments in spatial representations and rhythmic coordination of place cells in the 3xTg mouse model of Alzheimer's disease. \textit{Hippocampus}. \textbf{27}(4): 378-392 (2017).
	\bibitem{Canolty} Canolty, R., Edwards, E., Dalal, S., Soltani, M., Nagarajan, S., Kirsch, H., Berger, M., Barbaro, N., Knight, R. High Gamma Power Is Phase-Locked to Theta Oscillations in Human Neocortex. \textit{Science}. \textbf{313}(5793): 1626-8 (2006).
	\bibitem{Pastoll} Pastoll, H., Solanka, L., van Rossum, M., Nolan, M. Feedback inhibition enables $\theta$-nested $\gamma$ oscillations and grid firing fields. \textit{Neuron}. \textbf{77}(1): 141-54 (2013).
	\bibitem{Sirota} Sirota, A., Montgomery, S., Fujisawa, S., Isomura, Y., Zugaro, M, Buzs\'aki, G. Entrainment of neocortical neurons and gamma oscillations by the hippocampal theta rhythm. \textit{Neuron}. \textbf{60}(4): 683-97 (2008).
	
	%91-95
	\bibitem{CanoltyLearn} Canolty, R., Knight, R. The functional role of cross-frequency coupling. \textit{Trends Cog. Sci}. \textbf{14}(11): 506-15(2010). 
	\bibitem{Benchenane} Benchenane, K., Peyrache, A., Khamassi, M., Tierney, P., Gioanni, Y., Battaglia, F. \& Wiener, S. Coherent Theta Oscillations and Reorganization of Spike Timing in the Hippocampal- Prefrontal Network upon Learning. \textit{Neuron} \textbf{66}: 921-936 (2010).
	\bibitem{Zhang} Zhang, X., Zhong, W,, Brankack, J., Weyer, S., M\"uller, U., Tort, A., Draguhn, A. Impaired theta-gamma coupling in APP-deficient mice. \textit{Scientific reports}. \textbf{6}: 21948 (2016).
	\bibitem{Goodman} Goodman, M., Kumar, S., Zomorrodi, R., Ghazala, Z., Cheam, A., Barr, M., Daskalakis, Z., Blumberger, D., Fischer, C., Flint, A., Mah, L., Herrmann, N., Bowie, C., Mulsant, B., Rajji, T. Theta-Gamma Coupling and Working Memory in Alzheimer's Dementia and Mild Cognitive Impairment. \textit{Front Aging Neurosci}. \textbf{10}: 101 (2018). 
		
	\bibitem{BuzSharp} Buzs\'aki, G. Hippocampal sharp wave-ripple: A cognitive biomarker for episodic memory and planning. \textit{Hippocampus}. \textbf{25}(10): 1073-188 (2015).
	\bibitem{Csicsvari1} Csicsvari, J., Dupret, D. Sharp wave/ripple network oscillations and learning-associated hippocampal maps. \textit{Philos. Trans. R. Soc. B}. \textbf{369}(1635): (2014).
	
	%96-100
	\bibitem{Joo} Joo, H., Frank, L. The hippocampal sharp wave-ripple in memory retrieval for immediate use and consolidation. \textit{Nat. Rev. Neurosci}. \textbf{19}(12): 744-57 (2018).
	\bibitem{Leonard} Leonard, T., Mikkila, J., Eskandar, E., Gerrard, J., Kaping, D., Patel, S., Womelsdorf, T., Hoffman, K. Sharp Wave Ripples during Visual Exploration in the Primate Hippocampus. \textit{J Neurosci}. \textbf{35}(44): 14771-82 (2015).
	\bibitem{Fernandez} Fern\'andez-Ruiz, A., Oliva, A., Fermino de Oliveira, E., Rocha-Almeida, F., Tingley, D., Buzs\'aki, G. Long-duration hippocampal sharp wave ripples improve memory. \textit{Science}. \textbf{364}(6445): 1082-1086 (2019).
	\bibitem{Girardeau1} Girardeau, G., Benchenane, K., Wiener, S., Buzs\'aki, G. \& Zugaro, M. Selective suppression of hippocampal ripples impairs spatial memory. \textit{Nat. Neurosci}. \textbf{12}, 1222-1223 (2010).
	\bibitem{Schreiter} Schreiter-Gasser, U., Gasser, T., Ziegler, P. Quantitative EEG analysis in early onset Alzheimer's disease: correlations with severity, clinical characteristics, visual EEG and CCT. \textit{Clin. Neurophysiol}. \textit{90}(4): 267-272 (1994).
	
	%101-105
	\bibitem{Girardeau2} Girardeau, G., Zugaro, M. Hippocampal ripples and memory consolidation. \textit{Curr. Opin. Neurobiol} \textbf{21}, 452-459 (2001).
	\bibitem{Roux} Roux, L., Hu, B., Eichler, R., Stark, E. \& Buzs\'aki, G. Sharp wave ripples during learning stabilize the hippocampal spatial map. \textit{Nat. Neurosci}. \textbf{20}:845-853 (2017). %(6)
	\bibitem{Sadowski2} Sadowski, J., Jones, M. \& Mellor, J. Sharp-Wave Ripples Orchestrate the Induction of Synaptic Plasticity during Reactivation of Place Cell Firing Patterns in the Hippocampus. \textit{Cell reports} \textbf{14}, 1916-1929 (2016).
	\bibitem{Caccavano} Caccavano, A., Bozzelli, P. L., Forcelli, P. A., Pak, D. T. S., Wu, J. Y., Conant, K., Vicini, S. Inhibitory Parvalbumin Basket Cell Activity is Selectively Reduced during Hippocampal Sharp Wave Ripples in a Mouse Model of Familial Alzheimer's Disease. \textit{J Neurosci}. \textbf{40}(26): 5116–5136 (2020).
	\bibitem{Witton} Witton, J., Staniaszek, L., Bartsch, U., Randall, A., Jones, M., Brown, J. Disrupted hippocampal sharp-wave ripple-associated spike dynamics in a transgenic mouse model of dementia. \textit{J Physiol}. \textbf{594}(16): 4615-30 (2016). 
	
	%106-110
	\bibitem{Arenas1} Arenas, A., D\'az-Guilera, A., Kurths, J., Moreno, Y. \& Zhou, C., Synchronization in complex networks. \textit{Physics Reports} \textbf{469}: 93-153 (2008).
	\bibitem{Liao} Liao, X. et al. Pattern formation in oscillatory complex networks consisting of excitable nodes. \textit{Physical Review E} \textbf{83}: 056204 (2011).
	\bibitem{Mi} Mi, Y. et al. Long-period rhythmic synchronous firing in a scale-free network. \textit{Proc. Natl. Acad. Sci}. \textbf{110}: E4931-E4936 (2013).
	\bibitem{Restrepo} Restrepo, J., Ott, E. \& Hunt, B. Onset of synchronization in large networks of coupled oscillators. \textit{Physical Review E} \textbf{71}: 036151 (2005).
	
	%111-115
	\bibitem{Burton} Burton, S., Ermentrout, G. \& Urban, N. Intrinsic heterogeneity in oscillatory dynamics limits correlation-induced neural synchronization. \textit{J Neurophysiol} \textbf{108} (2012).
	\bibitem{Bezaire} Bezaire, M., Raikov, I., Burk, K., Vyas, D. \& Soltesz, I. Interneuronal mechanisms of hippocampal theta oscillations in a full-scale model of the rodent CA1 circuit. \textit{eLife}, \textbf{5}: e18566 (2016).
	\bibitem{BuzGamma} Buzs\'aki, G. \& Wang, X. Mechanisms of gamma oscillations. \textit{Ann. Rev. Neurosci} \textbf{35}: 203-225 (2012).
	
	\bibitem{Jun} Jun, H., Bramian, A., Soma, S., Saito, T., Saido, T., Igarashi, K. Disrupted Place Cell Remapping and Impaired Grid Cells in a Knock-in Model of Alzheimer's Disease. \textit{Neuron}. \textbf{107}(6): 1095-1112.e6 (2020).
	\bibitem{Cacucci2} Cacucci, F., Yi, M., Wills, T., Chapman, P., O'Keefe, J. Place cell firing correlates with memory deficits and amyloid plaque burden in Tg2576 Alzheimer mouse model. \textit{Proc. Natl. Acad. Sci}. \textbf{105}(22): 7863-8 (2008). 
	
	\bibitem{ArnB2} Arnold, V. Topology and statistics of arithmetic and algebraic formulae. \textit{Russian Math. Surv}. \textbf{58}(4) 637–664 (2003).
	\bibitem{ArnB4} Arnold, V. \textit{Lectures and Problems: A Gift to Young Mathematicians}, American Math Society, Providence (2015).
	\bibitem{Stephens} Stephens, M. Introduction to Kolmogorov (1933) On the Empirical Determination of a Distribution. In: Kotz S., Johnson N.L. (eds) Breakthroughs in Statistics. Springer Series in Statistics. Springer, New York, NY (1992). % (Perspectives in Statistics)
	\bibitem{Vrbik1} Vrbik, J. Small-sample corrections to Kolmogorov–Smirnov test statistic. \textit{Pioneer Journal of Theoretical and Applied Statistics}, \textbf{15}(1–2): 15–23 (2018).
	\bibitem{Bol1} Bol'shev, L. Asymptotically Pearson Transformations \textit{Theory Probab. Appl.}, \textbf{8}(2): 121-146 (1963).

\end{thebibliography}
\end{document}